\documentclass[aps,prd,superscriptaddress,notitlepage,11pt,final,longbibliography,nofootinbib]{revtex4-1}

\usepackage{color}

\usepackage[pdftex]{graphicx,xcolor}
\usepackage[pdftex]{hyperref}

\usepackage{float}

\usepackage{amsmath,amssymb,mathrsfs}

\usepackage{subfigure}
\usepackage{epsf}

\usepackage{bbm}
\usepackage{color}
\usepackage{comment}
\usepackage{cleveref}
\usepackage{comment}
\usepackage{natbib}
\usepackage{appendix}
\renewcommand{\bm}[1]{{\mbox{\boldmath $#1$}}}

\graphicspath{{./Fig/}}

\begin{document}
\title{Discovery of Quasi-Integrable Equations from traveling-wave data using the Physics-Informed Neural Networks}
	
\author{A. Nakamula~}
\email{nakamula@sci.kitasato-u.ac.jp}
\affiliation{Department of Physics, School of Science, Kitasato University, Sagamihara, Kanagawa 252-0373, Japan}

\author{K. Obuse~}
\email{obuse@okayama-u.ac.jp}
\affiliation{Graduate School of Natural Science and Technology, Okayama University, Okayama 700-8530, Japan}

\author{N. Sawado~}
\email{sawadoph@rs.tus.ac.jp}
\affiliation{Department of Physics and Astronomy, Tokyo University of Science, Noda, Chiba 278-8510, Japan}

\author{K. Shimasaki~}
\email{shimasakitus@gmail.com}
\affiliation{Department of Physics and Astronomy, Tokyo University of Science, Noda, Chiba 278-8510, Japan}

\author{Y. Shimazaki~}
\email{shimazakitus@gmail.com}
\affiliation{Department of Physics and Astronomy, Tokyo University of Science, Noda, Chiba 278-8510, Japan}

\author{Y. Suzuki~}
\email{ytszkyuta@gmail.com}
\affiliation{Department of Physics and Astronomy, Tokyo University of Science, Noda, Chiba 278-8510, Japan}

\author{K. Toda~}
\email{kouichi@yukawa.kyoto-u.ac.jp}
\affiliation{Department of Mathematical Physics, Toyama Prefectural University, Imizu, Toyama 939-0398, Japan}
\affiliation{Research and Education Center for Natural Sciences, Keio University, Hiyoshi 4-1-1, Yokohama, Kanagawa 223-8521, Japan}

\begin{abstract}
Physics-Informed Neural Networks (PINNs) have emerged as a powerful tool for analyzing nonlinear partial differential equations and identifying governing equations from observational data. In this study, we apply PINNs to investigate vortex-type solutions of quasi-integrable equations in two spatial dimensions, specifically the Zakharov-Kuznetsov (ZK) and the Regularized Long-Wave (RLW) equations. These equations are toy models for geostrophic shallow water dynamics in planetary atmospheres. We first demonstrate that PINNs can successfully solve these equations in the forward process using a mesh-free approach with automatic differentiation. However, in the inverse process, substantial misidentification occurs due to the structural similarities between the ZK and the RLW equations. To address this issue, we then introduce conservation law-enhanced PINNs, initial condition variations, and a friction-based perturbation approach to improve identification accuracy. Our results show that incorporating small perturbations while preserving conservation laws significantly enhances the resolution of equation identification. These findings may contribute to the broader goal of using deep learning techniques for discovering governing equations in complex fluid dynamical systems, such as Jupiter’s Great Red Spot.
\end{abstract}

\maketitle

\section{Introduction}

Non-linear wave phenomena widely appear in various physical systems with different scales, such as particle physics, condensed matter, biological problems, and planetary atmospheres. 
We believe that each phenomenon can be described by suitable partial differential equations (PDEs).   
Non-linear PDEs that admit soliton solutions are of special 
interest because of their mathematical beauty and consistency, i.e., their integrability.  
It supports properties such as the solutions having an infinite number of conservation laws, 
Painl\'eve properties, reducibility into bilinear forms. 
Probably, the most well-known equation with these properties is the Korteweg-de Vries (KdV) equation
\begin{align}
u_t+u_x+2uu_x+u_{xxx}=0, 
\label{KdV}
\end{align}
describing shallow water waves in 1+1 dimensions. 
Note that the second term of \eqref{KdV} is removed by taking $x'=x-t$ and $t$ as independent time
variables. Zabusky and Kruskal made the first observations of the basic property of the solitons in the equation~\cite{Zabusky-Kruskal}. 
On the other hand, Benjamin, Bona, and Mahony argued the superiority of the regularized long-wave (RLW) equation
\begin{align}
u_t+u_x+2uu_x-u_{xxt}=0
\end{align} 
over \eqref{KdV} for describing long-wave nature~\cite{Benjamin72}. 
They claimed these equations are known to be equivalent in the 
zeroth order, $u_t=-u_x$ of their perturbation.  
However, significant differences appear concerning these neighboring equation’s essential
mathematical and numerical aspects. 
Here, a natural question arises: if numerical data is given from an observation,
how do we identify the appropriate governing equation? 

Then, this paper concerns the identification of two 2+1 dimensional quasi-integrable equations, the Zakharov-Kuznetsov (ZK) equation and the RLW equation, which are supposed to be toy models of the shallow-water equations. The ZK equation given by~\cite{Zakharov74,IWASAKI1990293} 
\begin{align}
u_t+2uu_x+(\nabla^2u)_x=0
\end{align}
is a direct extension of the KdV equation into higher dimensions. 
To date, most of the previous research \cite{PetYan82, IWASAKI1990293, Klein21} are focused on the \(2 + 1\) dimensional version of the model. 
The equation possesses stable isolated vortex-type solutions 
which enjoy solitonic properties. It is well known that 
some inelastic properties emerge in a collision process when the strengths of vortices are significantly different, 
the taller soliton gains more height while the shorter one tends to wane with radiation~\cite{IWASAKI1990293}. 
The RLW equation~\cite{Peregrine66,Benjamin72} 
is a shallow-water fluid model to simulate the development of an undular bore. 
The RLW equation in \(2 + 1\) dimensions 
\begin{align}
u_t+u_x+uu_x-(\nabla^2u)_t=0
\end{align}
possesses a stable isolated vortex solution~\cite{KAWAHARA199279}. 
The peculiar inelastic properties of the solitons 
have been extensively studied in the literatures~\cite{ABDULLOEV1976427,COURTENAYLEWIS1979275,terBraak:2017jpe} in 1+1 dimensions 
and also~\cite{KAWAHARA199279} in 2+1 dimensions.
Stable fluid mechanical vortices in shallow water equations are often discussed in the context of quasi-integrable equations in 2+1 dimensions.
Notably, the significant dynamics of planetary atmospheres of Jupiter's or other planets 
are described by some non-linear PDEs in 2 or 3 spatial dimensions.
The Great Red Spot (GRS) of Jupiter is an illustration of a unique, extraordinary object because of its fantastic longevity. 
Many observations indicate that the phenomenon is undoubtedly shallow~\cite{Parisi21}. 
Its stable nature strongly suggests that there exists unknown mechanics coming from the highly non-linear nature 
of dynamics.
Models of shallow water equations and (or) their induced effective equations in several geostrophic regimes may describe the physics of the GRS. 
In the quasi-geostrophic regime, the Charney-Hasegawa-Mima (CHM) equation~\cite{Hasegawa78} and the drift-wave equation~\cite{Petviashvili80} by including the scalar non-linear term $uu_x$ 
\begin{align}
&(1-\nabla^2)u_t+u_x-J[u,\nabla^2 u]=0\,,
\\
&(1-\nabla^2)u_t+u_x+uu_x-J[u,\nabla^2 u]=0,
\label{driftwave}
\end{align}
respectively, are proposed, where the last term in equations, the Jacobian, is defined as 
$J[A,B]:= \left(\partial_x A\right) \left( \partial_y B\right) - \left(\partial_y A\right) \left(\partial_x B\right)$. 
These equations possess the vortex solutions with the dipole (CHM) and 
the monopole (drift-wave). 
In the intermediate geostrophic regime, the governing equation called
Williams-Yamagata-Flierl (WYF) equation~\cite{Williams84,charney1981oceanic,Koike:2022gfq}
(in cyclonic shear) is defined as
\begin{align}
&u_t+2uu_x+P(y)(\nabla^2 u)_x+2Q(y)u_x+2J[u,\nabla^2u]=0\,,
\\
&P(y)=1+2u^0(y),Q(y)=y+u^0_{yy}+\int^yu_0(y')dy',
\label{WYF}
\end{align}
where $u^0$ provides the effect of the background shear flow. 
With such various model equations in hand, the crucial step in solving a physical problem is to choose the appropriate governing equation. 
There are numerous potential possibilities for describing the GRS, 
but we are aware that the most appropriate equation for those phenomena is still unclear.
Discovery of the governing equation is thus crucial because it determines an 
appropriate physical regime for the phenomenon.  
At present, the observation data is still not enough to discover the equation for many phenomena including GRS. 
Thus, we do not go into realistic models (CHM, WYF), and focus on the toy models of these equations. 
Within a simple approximation $J\sim 0$~\footnote{For the solutions enjoying circular symmetry, the Jacobian is always zero. Note that our solutions in the present paper sometimes deviate from the circular shape.} 
the ZK and RLW equations can be toy models of the geostrophic equations. 
For the ZK equation, a shear flow $u^0=-1$~\cite{Koike:2022gfq} must be chosen.  
We believe that analysis of the identification in 
these equations will eventually be useful for determining geostrophic regimes of the above significant 
atmospheric problems.  

We study the identification of non-linear PDEs by means of deep learning technology. 
The application of deep learning in deep neural networks to non-linear PDEs has recently gained much attention. The Physics-Informed Neural Networks (PINNs) \cite{RaissarxivI,RaissarxivII,RAISSI2019686} 
have the ability to solve many complicated scientific problems, such as 
fluid and solid mechanics~\cite{Raissi2020,Brunton2020,Kadeethum2020,cai2021,Kashinath2021,Jin2021},
cyber-physical systems~\cite{9064519}, biological systems~\cite{Wu2017,KISSAS2020112623,Ruizherrera2021,SEL2023}, 
and many others.
Since the methods' original premise involved an application to fluid mechanics with PDEs~\cite{Raissi2020,Brunton2020} 
and a weather/climate modelling~\cite{Kashinath2021}, 
it is natural to extend the analysis into the significant dynamics of planetary atmospheres
described by some PDEs. 
One unique, noteworthy aspect of the PINNs is their ability to solve 
PDEs (the forward analysis) efficiently and also to provide us with an accurate estimate of the equation based on 
the governing data of the physical problems of our concern (the inverse analysis). 
PINNs generally have significantly higher extrapolation power than other conventional deep-learning techniques, 
making them appropriate for analyses involving limited learning data. 
In \cite{SIVALINGAM2024150,SIVALINGAM2024_2}, Sivalingam et al. have 
proposed sophisticated methods for solving several fractional-order PDEs.
Recently, there have been some attempts to apply the methods
to the more complex geometry of data~\cite{Fang2020,SAHLICOSTABAL2024107324}. 

In the present paper, we solve quasi-integrable equations via PINNs as reduced toy models
instead of directly solving above-mentioned geostrophic equations. 
The above geostrophic equations give us many insights into the phenomena, and the traditional approaches can, of course, work for the analysis of these equations. 
One advantage of solving the integrable equations or the quasi-integrable counterparts 
is to expose the corresponding phenomena and a primary character usually described by some mathematical language such as the existence of (infinite number of) conservation quantities. 
There are numerous studies of PINNs for the integrable equations, such as the 
Burgers equation~\cite{RaissarxivI,RaissarxivII,RAISSI2019686,LIN2022111053,JAGTAP2020113028}, 
the KdV equation and the modified KdV equtaion~\cite{Li_2020,JAGTAP2020113028,FANG2022112118,Junkai2024,Zhou2024,LIN2023133629}, 
the non-linear Schr\"odinger equation~\cite{ZHOU2021127010,Pu2021}, and many other variations. 
However, only few analyses have been studied in 2+1 dimensions.
There are the PINN analysis for 
the line solitons of the integrable Kadomtsev-Petviashvili
equation~\cite{Zhengwu2022} and also of the non-linear Schr\"odinger equation~\cite{ZHOU2023164,PENG2024}. 
The stationary vortex solutions in the nonlinear Schr\"odinger equation were studied in \cite{WANG202317}.
To our knowledge, no concrete research has yet been done on the ZK or the RLW equations. 
It should be noted that PINNs are distinct from the conventional, 
standard numerical algorithm~\cite{JAGTAP2020113028,MISHRA2021107705,Chen2022UsingPI,YANG2023109656,Sedykh_2024} 
that uses the finite difference and finite element methods, 
where the governing PDEs are eventually discretized over the computational domain.  
Providing a mesh-free approach is one of the main benefits of PINNs 
since automatic differentiation approximates the differential operators in the governing PDEs. 
Traditional numerical approaches strongly depend on grid settings, which may cause a curse of
dimensionality. On the other hand, PINN grid independence is undoubtedly efficient, particularly 
for solving high-dimensional problems and inverse analysis. 
A notable feature of these higher dimensional non-linear PDE is the quasi-integrable property, 
where a limited number of conservation quantities exist, or they tend to 
be broken in some situations such as during the inelastic collision~\cite{shimasaki24col}.  
Thus, it is worth using the PINNs that can realize conservation quantities. 
A benefit of utilizing PINNs is their ability to integrate the conservation laws of a system 
into analysis by incorporating the condition into the loss function. 
The method called cPINNs gains the accuracy of the analysis~\cite{JAGTAP2020113028,LIN2022111053,WU2022112143}. 

Data-driven discovery in the present paper is the inverse analysis of parameter inference of the governing equations. 
The procedure is as follows:(i)~We numerically solve the quasi-integrable equations using the standard algorithm
such as the Runge-Kutta method, or PINNs. (ii)~Using these data, we perform the PINN's inverse analysis to determine the coefficients of the equations and to see how the proper equation is identified.  
The PINN inverse analysis may result in substantial misidentification between the vortex solutions of the ZK and the RLW equations. That is, when the candidate equation has terms from both the ZK and the RLW equations, the inverse PINNs cannot determine the governing equation. 
These equations have a lot in common. For example, they have the same number of conserved quantities 
and possess the same form of traveling wave solution. It appears to be the origin of the substantial misidentification 
that could happen in PINN inverse analysis. 
The present paper provides several tentative resolutions for avoiding the problem.  

For the PDE discovery, a sparse learning algorithm~\cite{CHENZhao24,Thanasutives_2023,10401233} 
is quite efficient for getting proper information on the structure of the governing equation. 
The procedure is as follows. Firstly, one selects feasible candidates for the corresponding problem, and then we refine them to select the most appropriate equation using PINNs. 
With the equation we consider here because of their dimensionality, the degrees of freedom of the 
problem become large, thus carrying out the first part of the algorithms is very hard. 
Therefore, in the present paper, we concentrate on the second part, i.e., 
just the PINN analysis.

The present paper is based on the proceedings paper of \textit{the XII. International Symposium on Quantum Theory and Symmetries}~(QTS12)
~\cite{Nakamula_2023}. 
In the proceedings, we presented a few potential solutions to the problem, 
including variations of the initial conditions and a friction method in terms of the $y$-dependent sheared current, which will be thoroughly covered in the present paper.  
The approach is undoubtedly improved in this paper, and we obtain excellent convergent properties of the mean-squared error, which was not looked at in the previous study. 
In this paper, we will introduce a new friction that improves the identification ability compared to the method shown in the proceedings. We also focus on the cPINN analysis and the use of the two-soliton solutions for the inverse analysis, which are the inventive contents of the present paper.

The paper is organized as follows. In Section \ref{sec:2}, we give a brief introduction to the 
ZK and the RLW equations. The conservation 
quantities in these equations are derived in this section. 
We present an overview of PINNs in Section \ref{sec:3}. 
We demonstrate the misidentification of the solutions as well as the good performance of PINNs 
for the quasi-integrable equations.
We provide several (less effective) prescriptions to avoid the problem.  
In Section \ref{sec:4}, we give successful solutions introducing variations of the initial conditions and a small perturbation for the equations and showing how they work well. 
The conclusions and remarks are presented in the last section.

\section{\label{sec:2}$(2+1)$dimensional quasi-integrable systems and the stable vortices}

Though the term ``quasi-integrability'' is frequently employed in literature, 
it is still obscure mathematically. 
One nice, straightforward definition of a quasi-integrable system is 
a PDE that has only a finite number of exact conserved quantities\footnote{In \cite{Ferreira:2013nda,terBraak:2017jpe}, 
the authors give more thorough discussions about the concept of quasi-integrability. }. 
The quasi-integrable systems we investigate in this paper are based on the simple definition above. 
Their solutions resemble solitons in the integrable system and have just 
four conservation quantities. 
This paper investigates a well-known quasi-integrable system in two space dimensions 
that embody the underlying KdV dynamics for the stability of their vortex solutions.
The PINN analysis detects a signal resulting from partial conservation of the quantities, 
which will be seen in the next section.

\subsection{THe Zakharov-Kuznetsov equation}
The ZK equation 
\begin{equation}
\frac{\partial u}{\partial t}+2u\frac{\partial u}{\partial x}+\frac{\partial}{\partial x}\left(\nabla^2 u\right)=0, 
\label{ZK}
\end{equation}
originally was the model for the plasma dynamics in three dimensions 
with a uniform magnetic field~\cite{Zakharov74}. 
Since most of the later research~\cite{PetYan82, IWASAKI1990293, Klein21} 
have focused on the two-dimensional version of the model, the Laplacian specifies 
$\nabla^2=\partial_{x}^2+\partial_y^2$. 
Eq.(\ref{ZK}) possesses meta-stable isolated vortex-type solutions 
which enjoy the solitonic properties. The single soliton is dynamically stable, and the two solitons of almost 
same height appears to collide without merging or dissipating, similar to the well-known collision properties of the KdV solitons. This does not apply to the two solitons whose heights are significantly different. The taller soliton is amplified while the shorter one tends to wane with the radiation~\cite{IWASAKI1990293}.

The solutions of (\ref{ZK}) propagate in a specific direction with uniform speeds. 
Here, we set the direction in the positive $x$ orientation and velocity as $c$, namely assuming $u=U(\tilde{x}:=x-ct,y)$. 
Plugging it into (\ref{ZK}), we obtain
\begin{align}
\nabla^2U=cU-U^2\,,
\label{ZK circ}
\end{align}
where $\nabla^2=\partial_{\tilde{x}}^2+\partial_y^2$. 
A steady progressive exact wave solution is of the form
\begin{align}
U_{\rm 1d}=\frac{3c\,}{2}{\rm sech}^2\biggl[\frac{\sqrt{c}}{2}(\tilde{x}\cos\theta+y\sin\theta)\biggr],
\label{ZK wave}
\end{align}
where the $\theta$ is a given inclined angle of the solution. 
It indicates that this solution is just a trivial embedding of the KdV soliton into two spatial dimensions. 
There are several mathematical and computational achievements. 
On the other hand, \eqref{ZK circ} possesses another solution keeping circular symmetry \eqref{ZK wave}.
To find it,  we introduce cylindrical coordinates and rewrite the equation as 
\begin{align}
\frac{1}{r}\frac{d}{dr}\biggl(r\frac{dU(r)}{dr}\biggr)=c\,U(r)-{U}(r)^2\,,
\label{eq:ZKcylind}
\end{align}
where $r:=\sqrt{\tilde{x}^2+y^2}$. 
We can find the solutions with the boundary condition $U\to 0$ 
as $r\to \infty$ numerically, and the solutions form a one-parameter family of $c$ such as $U(r):=c\,F(\sqrt{c}\,r)$. 

The solutions of \eqref{ZK} exhibit soliton-like properties; however, they are not similar to the solitons in integrable systems such as those of the KdV equation, and the stability of the solutions may be supported by the infinite number of conserved quantities in the equation.  
Eq.\eqref{ZK} admits only the four integrals of motion~\cite{KUZNETSOV1986103}, which are
\begin{align}
&I_1:=\int i_1(y)dy= \int u dx dy,\ i_1(y):=\int u dx,
\label{ZKCQ1}
\\
&I_2:=\int\frac{1}{2}u^2dxdy\,,
\label{ZKCQ2}
\\
&I_3:=\int\biggl[\frac{1}{2}(\nabla u)^2-\frac{1}{3}u^3\biggr]dxdy\,,
\label{ZKCQ3}
\\
&\bm{I}_4:=\int \bm{r}u dxdy-t\bm{e}_x\int u^2dxdy\,,
\label{ZKCQ4}
\end{align}
where $\boldsymbol{r}$ and $\boldsymbol{e}_x$ are the two-dimensional position vector and the unit vector in the $x$-direction.
Here, $I_1$ is interpreted as the ``mass" of the solution, and $I_1(y)$ itself is conserved similarly to the case of the KdV equation. 
Also, $I_2, I_3, \bm{I_4}$ means the momentum, the energy, and the center of mass, respectively.  
We, therefore, conclude that the ZK equation is not an integrable system in the manner of ordinary soliton equations. 

There are several stability studies of the line solitons~\cite{YAMAZAKI20174336,KLEIN2023133722}, 
and particularly of the transverse instability in~\cite{Bridges2000,Johnson2010}.
For the other type of solitons such as the bell-shaped solitary waves with circular symmetry, 
the stability analysis was extensively studied in \cite{deBouard_1996}. 
Also, some transitions from the line soliton to the bell-shaped solitons were observed \cite{Frycz1989,Frycz1992}.
Some discussions, including the stability property, are also given in \cite{KUZNETSOV1986103}.

\subsection{The regularized long-wave equation}

The RLW equation was introduced by Peregrine~\cite{Peregrine66} to
simulate the development of the undular bore in the shallow water system, and  
many analytical and numerical studies have followed it such as \cite{Benjamin72, Goda1980, Medeiros1977, BHARDWAJ20001397,KAWAHARA199279,DEHGHAN20112540,DEHGHAN2015211}. 
The two-dimensional version of the RLW equation
\begin{equation}
\frac{\partial u}{\partial t}+\frac{\partial u}{\partial x}+u\frac{\partial u}{\partial x}
-\frac{\partial}{\partial t}\left(\nabla^2 u\right)=0 \label{RLW}
\end{equation}
concerns the drift waves in magnetized plasma or the Rossby waves in rotating fluids, 
where the field $u$ describes the mean horizontal velocity of the water. 
Our aim in the present paper is to mainly clarify the effects of the third-order 
derivatives terms $u_{xxx}$ in the ZK equation and $u_{xxt}$ in the RLW equation. 
For simplifying the discussion, instead of \eqref{RLW}, we employ 
\begin{align}
\frac{\partial u}{\partial t}+u\frac{\partial u}{\partial x}
-\frac{\partial}{\partial t}\left(\nabla^2 u\right)=0, \label{RLW_removed}
\end{align}
with removal of the term $u_x$ from \eqref{RLW}. 
It is merely a practical reason for the omission. 
As we will see in the next section, the misidentification problem occurs 
when we examine the inverse problem using the PINNs with the terms $u_{txx},u_{xxx}$ in the equations. 
Furthermore, adding $u_x$ to \eqref{ZK} and \eqref{RLW_removed} causes an additional identification problem that might happen between \(u_t\) and \(u_x\), which complicates the problem.  
It is straightforward to see that the modified equation \eqref{RLW_removed} also 
possesses the steady traveling solutions like \eqref{RLW}, which we shall see in the following. 
For the steady traveling solutions, the ZK and the RLW equations have similar fundamental properties. 
By plugging $u=U(\tilde{x},y)$ into (\ref{RLW}) or (\ref{RLW_removed}), we promptly get
\begin{align}
c\,\nabla^2U=(c-1)U-\frac{1}{2}U^2\,
\label{RLWt}
\end{align}
or
\begin{align}
c\,\nabla^2U=c\,U-\frac{1}{2}U^2\,.
\label{RLW_removedt}
\end{align}
The transformations of the fields leads
\begin{align}
&\textrm{the ZK~equation~\eqref{ZK}:}~~U=c\,F(X,Y),~~X:=\sqrt{c\,}\,\tilde{x},~~Y:=\sqrt{c\,}\,y\,,
(c>0)
\\
&\textrm{the RLW~equation~\eqref{RLW}:}~~U=2(c-1)F(X,Y),~~X:=\sqrt{1-\frac{1}{c}\,}\,\tilde{x},~~
Y:=\sqrt{1-\frac{1}{c}\,}\,y\,,
\nonumber \\
&\hspace{12cm}(c>1,c<0)
\nonumber \\
&\textrm{the modified RLW~equation~\eqref{RLW_removed}:}~~U=2c\,F(X,Y),~~X:=\tilde{x},~~Y:=y\,,
\end{align}
we promptly get a normalized identical equation
\begin{align}
\nabla^2F-F+F^2=0\,,~~\nabla^2:=\partial_X^2+\partial_Y^2\,.
\label{normalizedeq}
\end{align}
As we have already stated, \eqref{normalizedeq} possesses the traveling-wave solution
(like \eqref{ZK wave}) or the circular symmetric solution. 
Our central concern in the paper is that coincidence of transformed equations 
(\eqref{ZK}), (\eqref{RLW}) and (\eqref{RLW_removed}) often brings misidentification into the PINN analysis.
Another similarity between the two equations is that the modified RLW equation 
(we omit the term \textit{modified} in the following) also has 
the four conserved quantities, albeit with somewhat different forms
\begin{align}
&\bar{I}_1:=\int \bar{i}_1(y)dy= \int u dx dy,\ \bar{i}_1(y):=\int u dx,
\label{RLWCQ1}
\\
&\bar{I}_2:=\int\Bigl[(\nabla u)^2+u^2\Bigr]dxdy\,,
\label{RLWCQ2}
\\
&\bar{I}_3:=\int\biggl(\frac{1}{2}u^2+\frac{1}{6}u^3\biggr)dxdy\,,
\label{RLWCQ3}
\\
&\bar{\bm{I}}_4:
=\biggl[\bm{r}u+\bm{e}_x\frac{1}{2}\int_{t_0}^{t}(\nabla u)^2dt' \biggr]dxdy
-t\bm{e}_x\int\biggl\{u+\frac{1}{2}\Bigl[u^2+(\nabla u)^2\Bigr]\biggr\}dxdy\,.
\label{RLWCQ4}
\end{align}
The energy conservation is applied in $\bar{I}_2$, 
not the third quantity $I_3$ like the ZK equation. 
The stability discussions of the RLW-type equation have been given in \cite{Benjamin_sta}, and also~\cite{deBouard_1996}.

\section{\label{sec:3}Preliminary analysis}

The PINN's deep learning method is the most effective at solving initial-boundary problems (forward problems) and reconstructing equations (inverse problems), as it can find the governing equation from given data. The following introduces the PINN method of solving data-driven vortex solutions of the non-linear quasi-integrable \((2 + 1)\) dimensional differential equations.
Throughout analysis in the present paper, VS code is employed as the Integrated Development Environment for Python 
3.11.9, and Tensorflow 2.18.0 for building PINNs. For the optimization, SciPy 1.15.1 uses the L-BFGS-B.
The computation is performed on a 64-bit Windows system with an Intel(R) Core(TM)i9-14900KF CPU(3.2GHz)
and NVIDIA GeForce RTX 4090 GPUs. 

\begin{figure}[H]
	\centering
	\includegraphics[clip,width=14.0cm]{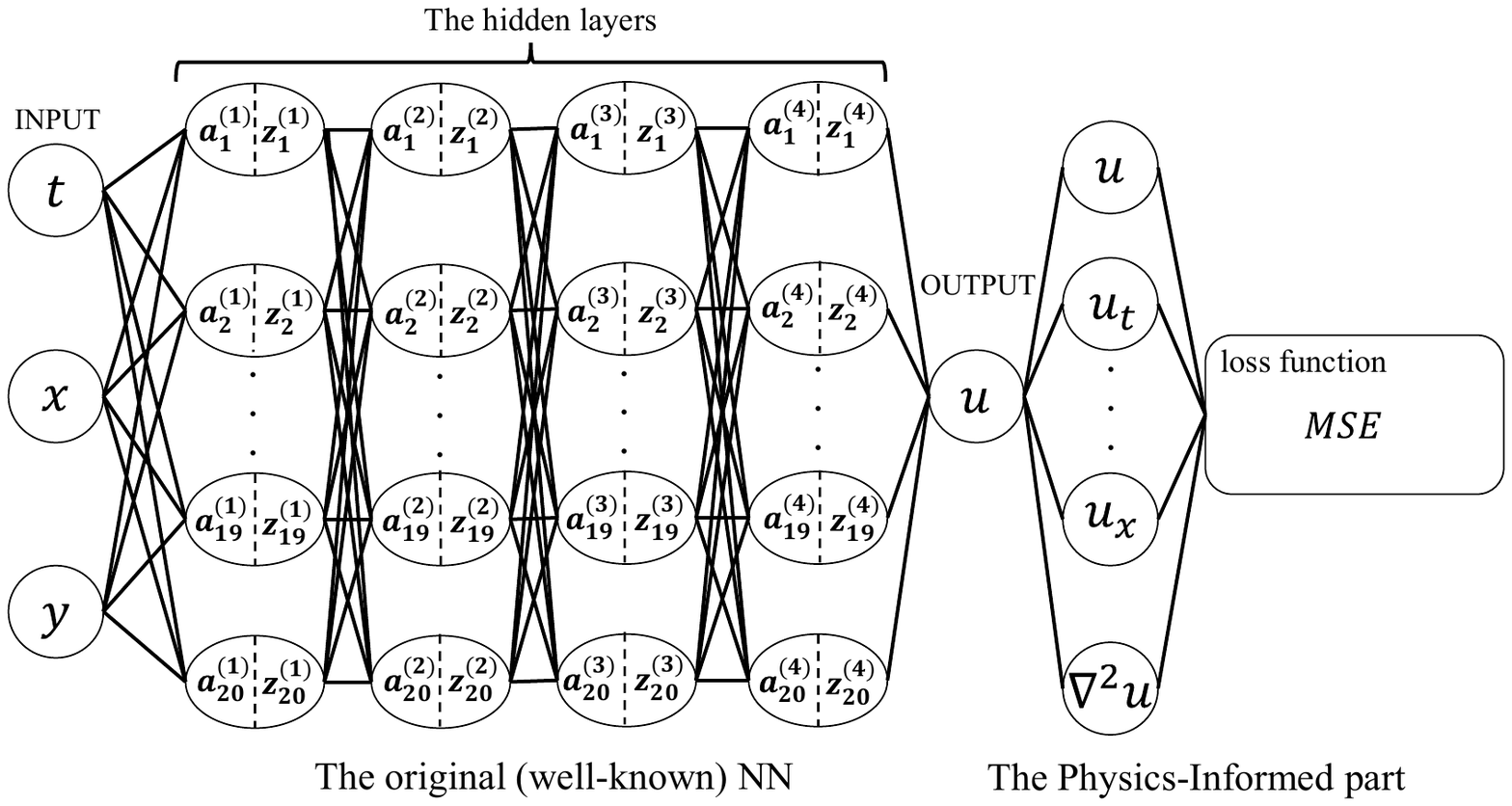}
	\caption{The PINNs: In the first part, we first calculate the weight sum, 
	$a^{(n)}_m=\sum_{i=1}^{20}\left\{w^{(n)}_{im}z^{(n-1)}_{i}+b^{(n)}_{m}\right\}$ 
	where $\left(w^{(j)}_{lk} , b^{(j)}_{k}\right)$ are the weighted parameters. 
	Second, we use $\tanh{x}$ as the activation function. 
	Finally, the output is obtained through the hidden layers. 
	After that, in the second part, we calculate the derivatives and construct a loss function.}
	\label{fig:pic1}
\end{figure}

\subsection{The PINNs for the (2+1)dimensional quasi-integrable systems}

The method consists of two segments of the networks (see figure 1). 
The first is the well-known neural network’s part, which is constructed with 4 hidden layers and 20 nodes for each layer to get the output $u$ from the inputs, and time-spatial coordinates $(x,y,t)$. 
However, the output of this network has no physical meaning. 
In the second segment, the output derivatives and a loss function for network optimization are estimated.

\begin{figure}[H]
	\centering 
	\includegraphics[width=1.0\linewidth]{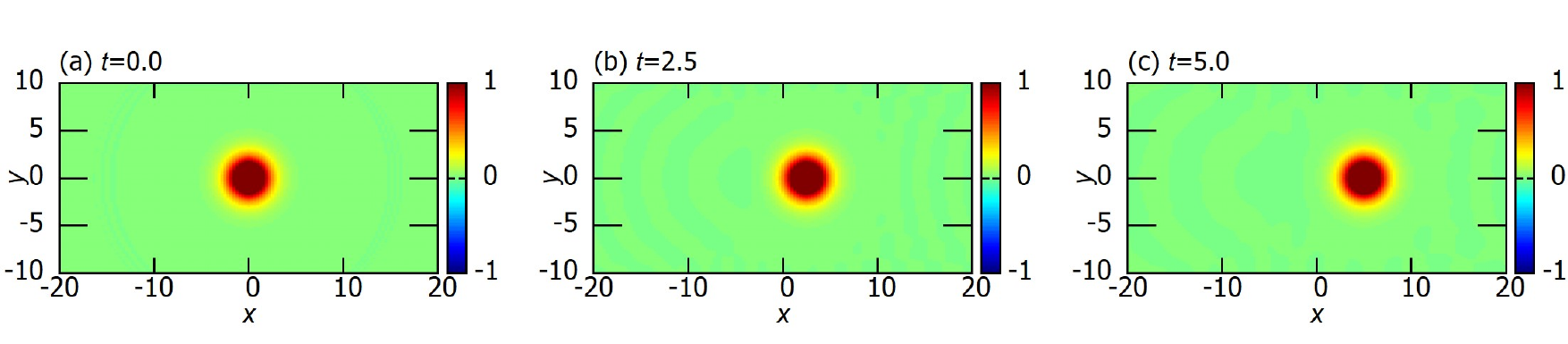}
	\caption{\label{ZK_fwd}The solution of the ZK equation in the forward PINN analysis.}

\end{figure}

\begin{figure}[H]
	\centering 
	\includegraphics[width=1.0\linewidth]{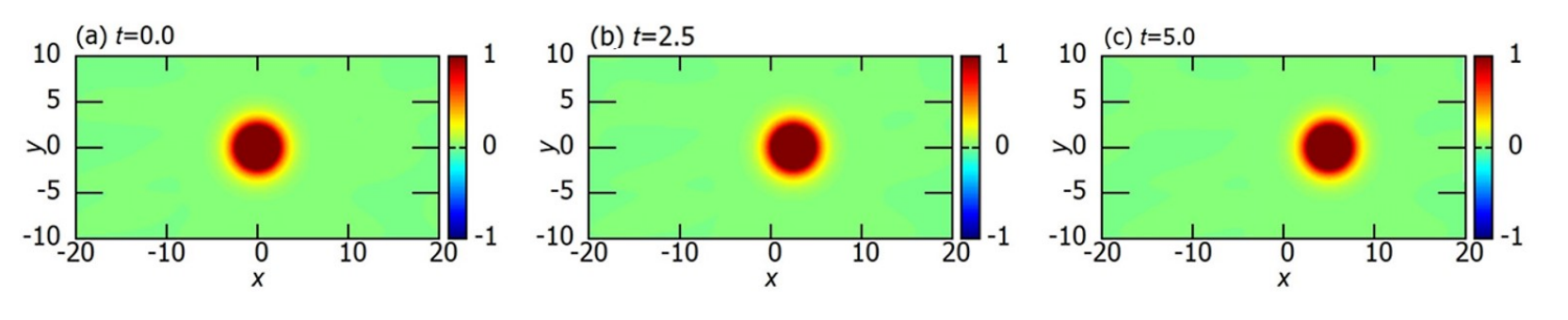}
	\caption{\label{RLW_fwd}The solution of the RLW equation in the forward PINN analysis.}
	
\end{figure}

The PINNs can be applied to both forward and inverse problems.  
The forward analysis is that one can solve a governing equation and find the solutions 
without ample computational resources or sophisticated numerical algorithms.  
Let us consider the PDEs
\begin{align}
&\textrm{ZK~eq.:}~~
\mathcal{F}_\textrm{ZK}:=u_t+\mathcal{N}_\textrm{ZK}(u,u_x,u_{xxx},u_{xyy})=0,
\\
&\textrm{RLW~eq.:}~~
\mathcal{F}_\textrm{RLW}:=u_t+\mathcal{N}_\textrm{RLW}(u,u_x,u_{txx},u_{tyy})=0,
\label{rlw0}
\end{align}
where $\mathcal{N}_\textrm{ZK},~\mathcal{N}_\textrm{RLW}$ are non-linear functions which are defined by
\begin{align}
\mathcal{N}_\textrm{ZK}(u,u_x,u_{xxx},u_{xyy}):=2uu_x+\left(\nabla^2 u\right)_x \label{zk}\,,
\\
\mathcal{N}_\textrm{RLW}(u,u_x,u_{txx},u_{tyy}):=2uu_x+\left(\nabla^2 u\right)_t \label{rlw}\,,
\end{align}
respectively. 
For simplicity, we have removed the term $u_x$ from the RLW equation. 
Although Eq.\eqref{rlw0} is not equivalent to the RLW equation in a strict sense, 
as the equation without the term \(u_x\) also reduces to the normalized equation~\eqref{normalizedeq}, a numerical traveling-wave solution still exists.
We focus on the vortex solutions moving to the positive $x$ direction of these equations.  
We define the rectangular mesh space 
\begin{align}
&x\in [-L_x,L_x],~N_x\textrm{th~grid~points};~~~~y\in [-L_y,L_y],~N_y\textrm{th~grid~points}\,,
\nonumber \\
&t\in [T_0,T_1]\,.
\end{align}
For optimization of the networks, the mean-squared error: $\mathit{MSE}$ is implemented as the loss function 
to measure the discrepancy between the predicted and the correct values, which
can be chosen such that
\begin{equation}
\mathit{MSE}=\frac{1}{N_u}\sum_{i=1}^{N_u}|u_{\mathrm{pred}}^{0}(x^i,y^i,0)-u_\mathrm{correct}^{0}(x^i,y^i,0)|^2
+\frac{1}{N_F}\sum_{i=1}^{N_F}|\mathcal{F}(x^i,y^i,t^i)|^2
+\frac{1}{N_\mathrm{b}}\sum_{N_{\mathrm{b}}}|\mathrm{BC} |^2,
\label{lossf}
\end{equation}
where $u^0_{\mathrm{pred}}$ is the predicted initial profile, $u^0_{\mathrm{correct}}$ is the correct initial profile, 
$\{x^i,y^i,0\}_{i=1}^{N_u}$ is the set of $N_u$th random residual points, and $\{x^i,y^i,t^i\}_{i=1}^{N_F}$ 
is the $N_F$th random points for the PINNs $\mathcal{F}(x,y,t)_\textrm{ZK}$ or $\mathcal{F}(x,y,t)_\textrm{RLW}$. 
Finally, $\mathrm{BC}$ represents the mean-squared error caused by the boundary conditions with the random $N_{\mathrm{b}}$th points. 
Here, the doubly periodic boundary condition is employed so that 
\begin{align}
\sum_{N_{\mathrm{b}}}|\textrm{BC} |^2
=\sum_{i=1}^{N_{\mathrm{b},x}}|u_{\mathrm{pred}}(x^i, L_y,t^i)-u_{\mathrm{pred}}(x^i,-L_y,t^i)|^2
\nonumber \\
+\sum_{j=1}^{N_{\mathrm{b},y}}|u_{\mathrm{pred}}(L_x,y^j,t^j)-u_{\mathrm{pred}}(-L_x,y^j,t^j)|^2\,,
\label{lossbc}
\end{align}
where $N_{\mathrm{b}}\equiv N_{\mathrm{b},x}+N_{\mathrm{b},y}$. For optimizing  the loss function, 
we employ the L-BFGS-B method; one of the quasi-Newton methods that can efficiently accelerate the convergence of the \textit{MSE}.
When the \textit{MSE} asymptotes to 0, the output $u$ reaches the exact solution of the PDE. In figures \ref{ZK_fwd}
and \ref{RLW_fwd}, we present the typical solutions of the ZK equation and the RLW equation in terms of the forward analysis.

Another prominent application of PINNs is the guessing of the PDE from provided data.
Now, we introduce a slightly modified PINNs
\begin{align}
&\textrm{ZK~eq:}~~
\tilde{\mathcal{F}}_\textrm{ZK}:=u_t+\tilde{\mathcal{N}}_\textrm{ZK}(u,u_x,u_{xxx},u_{xyy},\bm{\lambda})=0,
\\
&\textrm{RLW~eq:}~~
\tilde{\mathcal{F}}_\textrm{RLW}:=u_t+\tilde{\mathcal{N}}_\textrm{RLW}(u,u_x,u_{txx},u_{tyy},\bm{\lambda})=0,
\end{align}
where the unknown parameters $\bm{\lambda}$ is implemented in the equations. For the inverse analysis, 
we employ the different \textit{MSE} from \eqref{lossf},\eqref{lossbc}
\begin{equation}
\mathit{MSE}=\frac{1}{N_u}\sum_{i=1}^{N_u}|u_{\mathrm{pred}}(x^i,y^i,t^i)-u_\mathrm{correct}(x^i,y^i,t^i)|^2
+\frac{1}{N_F}\sum_{i=1}^{N_F}|\tilde{\mathcal{F}}(x^i,y^i,t^i)|^2, 
\label{lossi}
\end{equation}
where $u_{\mathrm{pred}}$ is the predicted profile, $u_\mathrm{correct}$ is the correct profile including the boundary data, $\{x^i,y^i,t^i\}_{i=1}^{N_u}$ is the set of $N_u$th random residual points, and $\{x^i,y^i,t^i\}_{i=1}^{N_F}$ is the $N_F$th random points for the PINNs $\tilde{\mathcal{F}}(x,y,t)_\textrm{ZK}$ or $\tilde{\mathcal{F}}(x,y,t)_\textrm{RLW}$.
Automatically choosing the parameters of the Neural Networks $\left(w^{(j)}_{lk} , b^{(j)}_{k}\right)$ 
and also the parameters $\bm{\lambda}$, the networks are properly optimized. 
As a result, we obtain the proper $\bm{\lambda}$ and then the idealized equations
for corresponding training data.

\begin{table}[H]
	\centering
	\caption{The successful parameterization for the ZK equation.}
	\label{tab:Inv_ZK}
	\centering
	\begin{tabular}{|c|c|c|}
		\hline
		& $\mathit{PDE}$ & $\mathit{MSE}$ $(\times 10^{-6})$ 
		\\ \hline
		Correct equation & $u_t+2uu_x+\left(\nabla^2 u\right)_x=0$ & $-$ 
		\\ \hline
		Identified (1) &$u_t+2.046uu_x+1.0048\left(\nabla^2 u\right)_x=0$ & 4.4 
		\\ \hline
		Identified (2) &$u_t+2.011uu_x+1.0033\left(\nabla^2 u\right)_x=0$ & 4.6
		\\ \hline
	\end{tabular}
\end{table} 

\begin{table}[H]
	\centering
	\caption{The successful parameterization for the RLW equation.}
	\label{tab:Inv_RLW}
	\centering
	\begin{tabular}{|c|c|c|}
		\hline
		& $\mathit{PDE}$ & $\mathit{MSE}$ $(\times 10^{-6})$ 
		\\ \hline
		Correct equation & $u_t+uu_x-\left(\nabla^2 u\right)_t=0$ & $-$ 
		\\ \hline
		Identified (1) &$u_t+0.9999uu_x-0.9996\left(\nabla^2 u\right)_t=0$ & 1.3 
		\\ \hline
		Identified (2) &$u_t+0.9989uu_x-0.9981\left(\nabla^2 u\right)_t=0$ & 1.4
		\\ \hline
	\end{tabular}
\end{table} 
\subsection{Successful analyses for the ZK equation and the RLW equation}

As a demonstration, we employ the PINNs to solve evolution equations and the solutions 
found in the profile $u$. 
We define $\tilde{\mathcal{N}}$ for the inverse analysis for the ZK equation, 
using two unknown constants $\lambda_0,\lambda_1$
\begin{equation}
\tilde{\mathcal{N}}_\textrm{ZK}(u,u_x,u_{xxx},u_{xyy};\lambda_0,\lambda_1):=\lambda_0uu_x+\lambda_1\left(\nabla^2 u\right)_x\,. \label{zk_id}
\end{equation}
We investigate the inverse analysis of the PINNs for finding the parameter values ($\lambda_0,\lambda_1$) 
using the training data derived from the forward analysis. In the PINN analysis, 
the numerical uncertainty is an inevitable consequence of training data sampling.  
We establish the correctness of our research in this paper by using the prescription of doing double identification for the same training data. In Table \ref{tab:Inv_ZK}, we present the result of the 
inverse analysis for the Zakharov-Kuznetsov equation. One can see that  
the PINN method reproduces the governing equation from the training data.
For the RLW equation, the PINN $\tilde{\mathcal{N}}$ is defined as
\begin{align}
\tilde{\mathcal{N}}_\textrm{RLW}(u,u_x,u_{txx},u_{tyy},\lambda_0,\lambda_1):=2\lambda_0 uu_x+\lambda_1\left(\nabla^2 u\right)_t\,. \label{rlw_id}
\end{align}
In the following, we shall omit the term $u_x$ in the equation for simplicity. 
The result of the PINN inverse analysis is presented in Table \ref{tab:Inv_RLW}.

The PINN functions $\tilde{\mathcal{N}}_\textrm{ZK}$ (\ref{zk_id}), $\tilde{\mathcal{N}}_\textrm{RLW}$ (\ref{rlw_id}) considered here are completely the same as the function of the forward analysis $\mathcal{N}_\textrm{ZK}$ (\ref{zk}), $\mathcal{N}_\textrm{RLW}$ (\ref{rlw}), respectively making the PINN analysis successful, as shown here. 
Such excellent predictive power is often lost when we try to introduce a more general variant of the \eqref{zk_id}  for the inverse analysis, \textit{i.e.}, to incorporate more complex nonlinear terms that do not exist in the original equations. 
The next subsection focuses on the misidentification between traveling wave solutions of the ZK equation and the RLW equation.

\begin{table}[t]
	\centering
		\caption{The misidentification for the ZK equation.}
		\label{tab:misidentify_ZK}
		\begin{tabular}{|c|c|c|}
			\hline
		& $\mathit{PDE}$ & $\mathit{MSE}$ $(\times 10^{-6})$
		\\ \hline
		Correct equation & $u_t+2uu_x+0.0\left(\nabla^2 u\right)_t+\left(\nabla^2 u\right)_x=0$ & $-$
		\\ \hline
		Identified $c=2$(1) &$u_t+2.0011uu_x-0.4017\left(\nabla^2 u\right)_t+0.2019\left(\nabla^2 u\right)_x=0$ & 3.8
		\\ \hline
		Identified $c=2$(2) &$u_t+2.0015uu_x-0.4796\left(\nabla^2 u\right)_t+0.0116\left(\nabla^2 u\right)_x=0$ & 4.0
		\\ \hline
		Identified $c=1$(3) &$u_t+2.0048uu_x-0.5190\left(\nabla^2 u\right)_t+0.5311\left(\nabla^2 u\right)_x=0$ & 4.1
		\\ \hline
		Identified $c=1$(4) &$u_t+2.0452uu_x-0.61035\left(\nabla^2 u\right)_t+0.4358\left(\nabla^2 u\right)_x=0$ & 5.2
		\\ \hline
		\end{tabular}
	\end{table}

\begin{table}[H]
	\centering
		\caption{The misidentification for the RLW equation.}
		\label{tab:misidentify_RLW}
		\begin{tabular}{|c|c|c|}
			\hline
			& $\mathit{PDE}$ & $\mathit{MSE}$ $(\times 10^{-6})$
			\\ \hline
		Correct equation & $u_t+uu_x-\left(\nabla^2 u\right)_t+0.0\left(\nabla^2 u\right)_x=0$ & $-$ 
		\\ \hline
		Identified $c=2$(1) &$u_t+0.9997uu_x-2.0549\left(\nabla^2 u\right)_t-2.1105\left(\nabla^2 u\right)_x=0$ & 5.2
		\\ \hline
		Identified $c=2$(2) &$u_t+0.9997uu_x-1.0784\left(\nabla^2 u\right)_t-0.1579\left(\nabla^2 u\right)_x=0$ & 4.8
		\\ \hline
		Identified $c=1$(3) &$u_t+0.9998uu_x-0.8330\left(\nabla^2 u\right)_t+0.1666\left(\nabla^2 u\right)_x=0$ & 3.8
		\\ \hline
		Identified $c=1$(4) &$u_t+0.9988uu_x-0.9952\left(\nabla^2 u\right)_t+0.0030\left(\nabla^2 u\right)_x=0$ & 5.9
		\\ \hline
		\end{tabular}
	\end{table}

\subsection{The misidentification for the ZK equation and the RLW equation}

We consider the following equation
\begin{equation}
u_t+\lambda_0uu_x+\lambda_1\left(\nabla^2 u\right)_t+\lambda_2\left(\nabla^2 u\right)_x=0.
\label{equation:ZKRLW}
\end{equation}
When we choose $(\lambda_0,\lambda_1,\lambda_2)=(2.0,0.0,1.0)$, \eqref{equation:ZKRLW} becomes 
the ZK equation, and for $(\lambda_0,\lambda_1,\lambda_2)=(1.0,-1.0,0.0)$, it reduces to the RLW equation. 
We study the inverse analysis both for the 1-soliton data of the ZK and the RLW equations.
The PINN fails to distinguish the equations: so-called \textit{misidentification}. 
We verify the inverse analysis that for the data obtained \eqref{equation:ZKRLW} leads to the ZK equation for $(\lambda_0,\lambda_1,\lambda_2)=(2.0,0.0,1.0)$ 
and to the RLW equation for $(\lambda_0,\lambda_1,\lambda_2)=(1.0,-1.0,0.0)$. 
As mentioned in Sec.II, the PINNs cannot distinguish these equations because
they possess essentially the same traveling-wave solution $u(x,y,t)=U(x-ct,y)$.
In Table \ref{tab:misidentify_ZK}, we present the result of the identification of the ZK equation with the training data of the ZK vortex
solution. Also, Table \ref{tab:misidentify_RLW} shows the identification of the RLW equation with the training data of the RLW vortex. 
Both identifications are completely failed.

In the following, we briefly look more closely at the origin of the stuck. 
Plugging the traveling-wave solution into (\ref{equation:ZKRLW}), we obtain 
\begin{equation}
-cU+\frac{\lambda_0}{2}U^2+\left(\lambda_2-c\lambda_1\right)\nabla^2 U=0. \label{static}
\end{equation}
If \eqref{static} comes from the ZK equation \eqref{ZK}, 
the coefficients should be
\begin{equation}
\lambda_0=2.0,~~
\lambda_2-c\lambda_1=1.0, 
\end{equation}
while from the RLW equation \eqref{RLW} with $u_x=0$, it is
\begin{equation}
\lambda_0=1.0,~~
\lambda_2-c\lambda_1=c.
\end{equation}
Consequently, $\lambda_1$ and $\lambda_2$ cannot be uniquely determined. 
The reason for the problem is straightforward:
Since the $t$ derivative and the $x$ derivative for the traveling waves are 
convertible with one another, misinterpretation of the equation 
inevitably would arise in the PINN analysis. If one wishes to use the PINNs efficiently, 
such coincidence must be suitably broken. 

\subsection{The conservative PINNs: The conservation constrained deep learning}

Including the conserved quantities of the ZK equation~\eqref{ZKCQ1}-\eqref{ZKCQ4} and
the RLW equation~\eqref{RLWCQ1}-\eqref{RLWCQ4} 
into the \textit{MSE} could potentially improve the accuracy of the guessing original equations.
Here, we employ the second conserved quantity of the ZK equation
\begin{align}
I_2(t)=\int \frac{u^2}{2}dxdy\,,
\end{align}
which is not the conserved quantity of the RLW equation because
\begin{equation}
\frac{dI_2(t)}{dt} = \int \left(u_xu_{xt}+u_yu_{yt}\right)dxdy 
\label{di2}
\end{equation}
is not zero when we use \eqref{RLW} into the righthand side of \eqref{di2}. 

\begin{figure}[h]
	\begin{center}
		\includegraphics[width=0.7\linewidth]{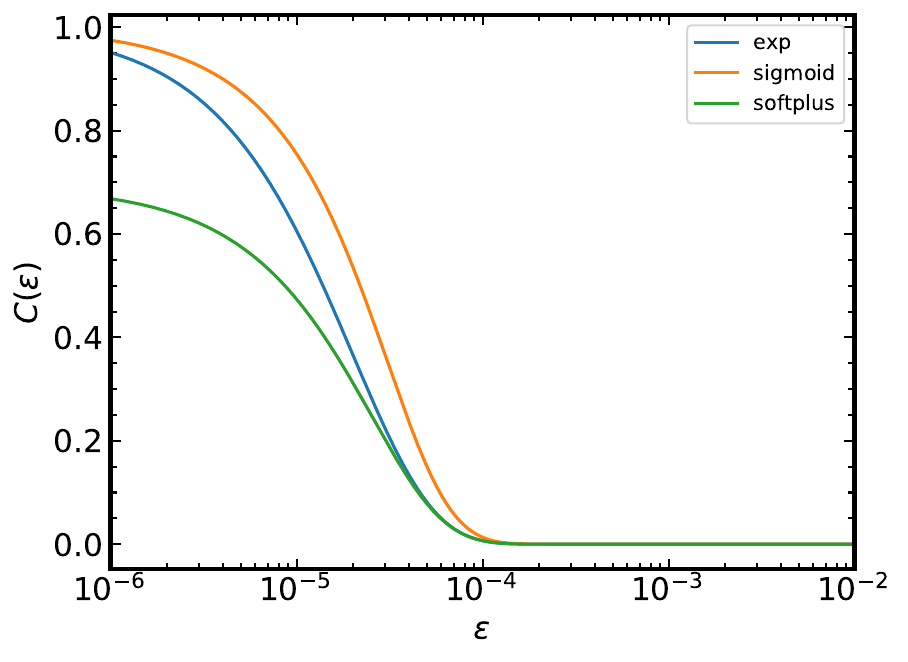}\\
	\end{center}
	\caption{\label{coefc}The behavior of $C(\mathcal{E})$ defined by \eqref{weitc},\eqref{weitf} 
	for $\gamma= 5.0\times10^{4}$ and $\mathcal{E}_\textrm{crit}=1.0\times10^{-3}$.}
\end{figure}

We define the cPINNs: the conserved PINN algorithm which is realized by the following \textit{MSE}
\begin{align}
MSE:=\mathcal{E}+C(\mathcal{E})\sum_{j=1}^{N_c}|I_2^{\textrm{pred}}(t^j)-I_2^{\textrm{correct}}|^2\,,
\label{cPINNloss}
\end{align}
where the $\mathcal{E}$ is identical to the standard $MSE$ that was previously introduced for the forward problem~\eqref{lossf}
\begin{align}
\mathcal{E}=\frac{1}{N_u}\sum_{i=1}^{N_u}|u_{\mathrm{pred}}^{0}(x^i,y^i,0)-u_\mathrm{correct}^{0}(x^i,y^i,0)|^2
+\frac{1}{N_F}\sum_{i=1}^{N_F}|\mathcal{F}(x^i,y^i,t^i)|^2
+\frac{1}{N_\mathrm{b}}\sum_{N_{\mathrm{b}}}|\mathrm{BC} |^2\,,
\end{align}
and for the inverse problem~\eqref{lossi}
\begin{align}
\mathcal{E}=\frac{1}{N_u}\sum_{i=1}^{N_u}|u_{\mathrm{pred}}(x^i,y^i,t^i)-u_\mathrm{correct}(x^i,y^i,0)|^2
+\frac{1}{N_F}\sum_{i=1}^{N_F}|\tilde{\mathcal{F}}(x^i,y^i,t^i)|^2\,.
\end{align} 
The $N_c$ is the number of the reference time for evaluating the conserved quantity $I_2^{\textrm{pred}}(t^j)$; here,
we evaluate the integration at $t^j=0.0,1.0,2.0,3.0,4.0$ (i.e., $N_c=5$) using the standard Simpson's method
\footnote{More sophisticated formula such as the Gauss quadrature 
is available but no notable difference occurs in the final result.}. 
$I_2^{\textrm{correct}}$ is the correct data of the conservation quantity, which is calculated by integrating $u_\mathrm{correct}(x^i,y^i,0)$.
Since the numerical integrals of the conserved quantity diverge from the true values for non-converged solutions, which leads the undesirable loss behavior, the basic idea of the cPINNs is that the $\mathcal{E}$ is trained first before the conservation law is considered. 
Now we discuss the dependence of the selection of the weight function $C(\mathcal{E})$ on the effectiveness of our study. 
The weight function $C(\mathcal{E})$ is defined in terms of a parameter $\gamma$ and a critical value of $\mathcal{E}_\textrm{crit}$, such that
\begin{align}
C(\mathcal{E})=\left\{
\begin{aligned}
&f(\mathcal{E}_\textrm{crit})~~~\mathcal{E}> \mathcal{E}_\textrm{crit}  \\
& f(\mathcal{E})~~~~~~\mathcal{E}\le \mathcal{E}_\textrm{crit} \, ,
\end{aligned}
\right.
\label{weitc}
\end{align}
where $f$ is an analytic function realizing the better resolution of PINNs. 
In the present paper, we show the results using the following three types of \(f(\mathcal{E})\)
\begin{align}
f(\mathcal{E})=
\left\{
\begin{array}{ll}
\exp(-\gamma\mathcal{E}) & \textrm{:exponential}
\\
\dfrac{2}{1+\exp(\gamma\mathcal{E})} & \textrm{:sigmoid}
\\
\log\left(1+\exp(-\gamma\mathcal{E})\right) & \textrm{:softplus}\,.~~
\end{array}
\right.
\label{weitf}
\end{align}
We choose the parameters as $\gamma=5.0\times10^{4},\mathcal{E}_\textrm{crit}=1.0\times10^{-3}$ of $C(\mathcal{E})$ 
by trial and error, realizing the lowest value of \textit{MSE}s.
We plot the behavior of \(C(\mathcal{E})\) in Fig.\ref{coefc}.
We demonstrate the identification of the ZK equation with three types of the 
weight functions \eqref{weitf}. 
One can easily find that the results do not significantly depend on the choice of the weight functions
shown in Table \ref{c_test}. 
In the following, we only employ the exponential function and change the value of the coupling parameter $\gamma$.

\begin{table}[H]
	\centering
	\caption{\label{c_test}The identification of the ZK equation in the \textit{MSE}~(\ref{cPINNloss}). 
	The parameter of $f(\mathcal{E})$ is fixed as $\gamma=5.0\times 10^4$.}
	\centering
	\begin{tabular}{|c|c|c|c|}
		\hline
		& $f(\mathcal{E})$ & $\mathit{PDE}$&$\mathit{MSE}$ $(\times 10^{-6})$\\ \hline
		\textrm{Correct equation}~&- &$u_t+2uu_x+\left(\nabla^2 u\right)_x=0$& $-$ \\ \hline
		\textrm{Identified} & exp& $u_t+2.019uu_x+0.5659\left(\nabla^2 u\right)_x-0.4524\left(\nabla^2 u\right)_t=0$&2.0\\ \hline
		\textrm{Identified} & sigmoid& $u_t+1.920uu_x+0.7121\left(\nabla^2 u\right)_x -0.5641\left(\nabla^2 u\right)_t=0$&11\\ \hline
		\textrm{Identified} & softplus& $u_t+1.751uu_x+0.3289\left(\nabla^2 u\right)_x-0.3316\left(\nabla^2 u\right)_t=0$&35\\ \hline
	\end{tabular}
\end{table}

\begin{figure}[h]
	\begin{center}
		\includegraphics[width=1.0\linewidth]{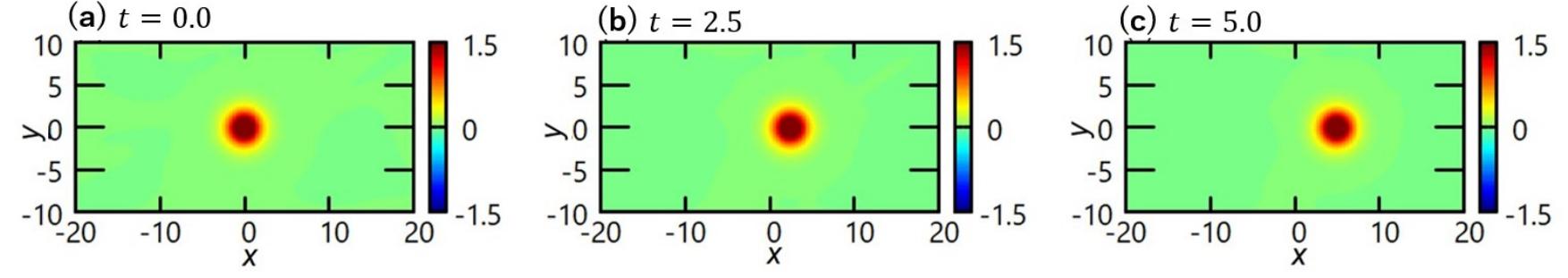}\\
	\end{center}
	\caption{\label{cPINN} single soliton profile predicted by cPINNs. the parameters for the $C(\mathcal{E})$ 
	are fixed as $\gamma= 5.0\times10^{4}$,  $\mathcal{E}_\textrm{crit}=1.0\times10^{-3}$.}
\end{figure}

\begin{figure}[h]
	\begin{center}
		\includegraphics[width=1.0\linewidth]{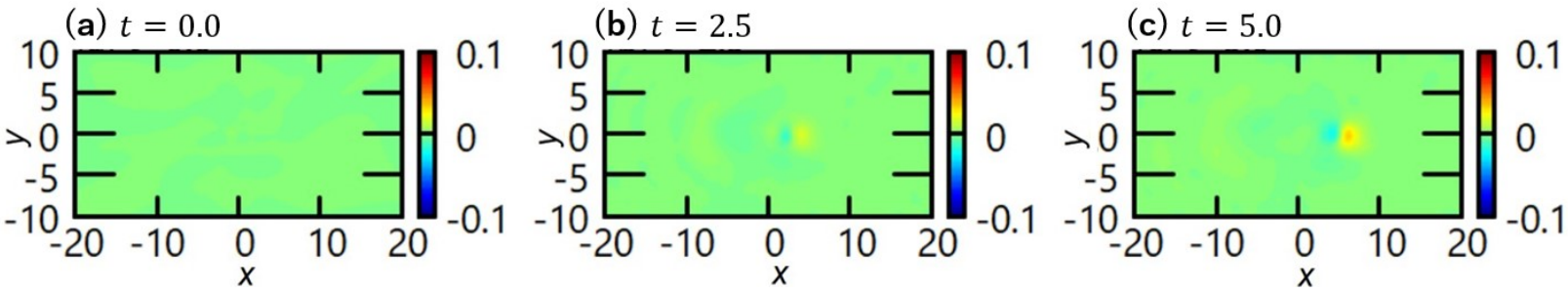}\\
		\textrm{(a)}~ cPINNs \\\
		\includegraphics[width=1.0\linewidth]{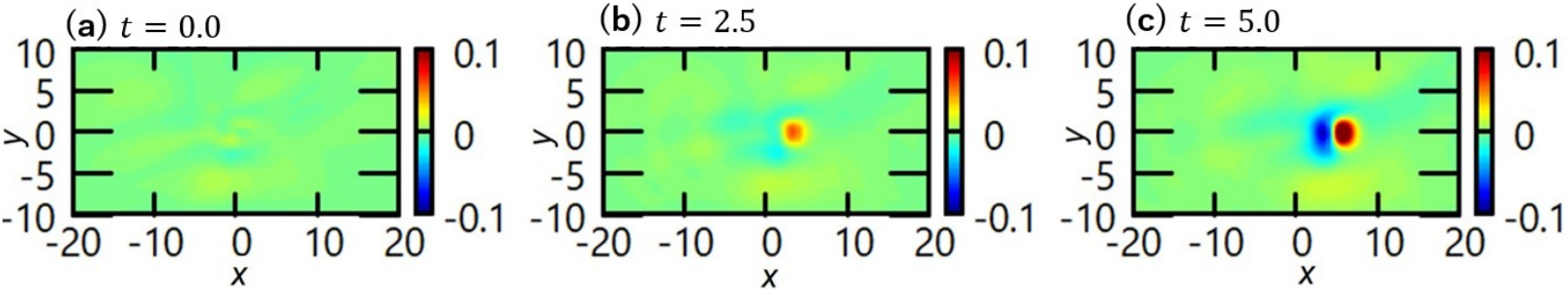}\\
		\textrm{(b)}~PINNs \\
	\end{center}
	\caption{\label{cPINNvsPINN} The error between the predicted profile and the numerical solution:
	 $u_{\textrm{pred}}-u_{\textrm{correct}}$ of the (a)~cPINNs, and (b)~PINNs.}
\end{figure}

The profile of the single soliton via cPINNs is presented in Fig.\ref{cPINN}.  
To verify the accuracy of the cPINNs, we assess the variation from the numerical analysis:~$u_\textrm{pred}-u_\textrm{correct}$, 
and the predictability of the cPINNs is much better than that of the PINN analysis as we can see in Fig.\ref{cPINNvsPINN}. 

However, as shown in Table \ref{zk_id_cPINN}, for the predictability compare with the PINNs, 
cPINNs are ineffective.  
The main reason for the failure is that the cPINN's projected profile is more accurate than that of PINNs and 
the output of the cPINNs is closer to $u_{\mathrm{pred}}(x,y,t)\sim U(x-ct,y)$. 
Unfortunately, the misidentification may be the result of our attempt to improve the predictability 
by considering the conserved quantities as described in this section.

\begin{table}[H]
	\centering
	\caption{\label{zk_id_cPINN}The identification of the ZK equation in the \textit{MSE}~(\ref{cPINNloss}) with 
	changing the coupling parameter $\alpha$.}
	\label{tab:Inv_ZK_cPINN}
	\centering
	\begin{tabular}{|c|c|c|c|}
		\hline
		& $\gamma~(\times 10^{3})$& $\mathit{PDE}$ & $\mathit{MSE}$ $(\times 10^{-6})$\\ \hline
		\textrm{Correct equation}~& $-$ &$u_t+2uu_x+\left(\nabla^2 u\right)_x=0$& $-$ \\ \hline
		\textrm{Identified} & 50 & $u_t+2.019uu_x+0.5659\left(\nabla^2 u\right)_x-0.4524\left(\nabla^2 u\right)_t=0$&2.0\\ \hline
		\textrm{Identified} & 10 & $u_t+2.020uu_x+0.5648\left(\nabla^2 u\right)_x -0.4524\left(\nabla^2 u\right)_t=0$&2.0\\ \hline
		\textrm{Identified} & 5 & $u_t+2.020uu_x+0.5676\left(\nabla^2 u\right)_x-0.4511\left(\nabla^2 u\right)_t=0$&2.2\\ \hline
		\textrm{Identified} & 1 & $u_t+0.6362uu_x+0.0351\left(\nabla^2 u\right)_x-0.00913\left(\nabla^2 u\right)_t=0$&611\\ \hline
	\end{tabular}
\end{table}

\subsection{The 2-soliton solution}

\begin{figure}[H]
	\centering
	\includegraphics[width=1.0\linewidth]{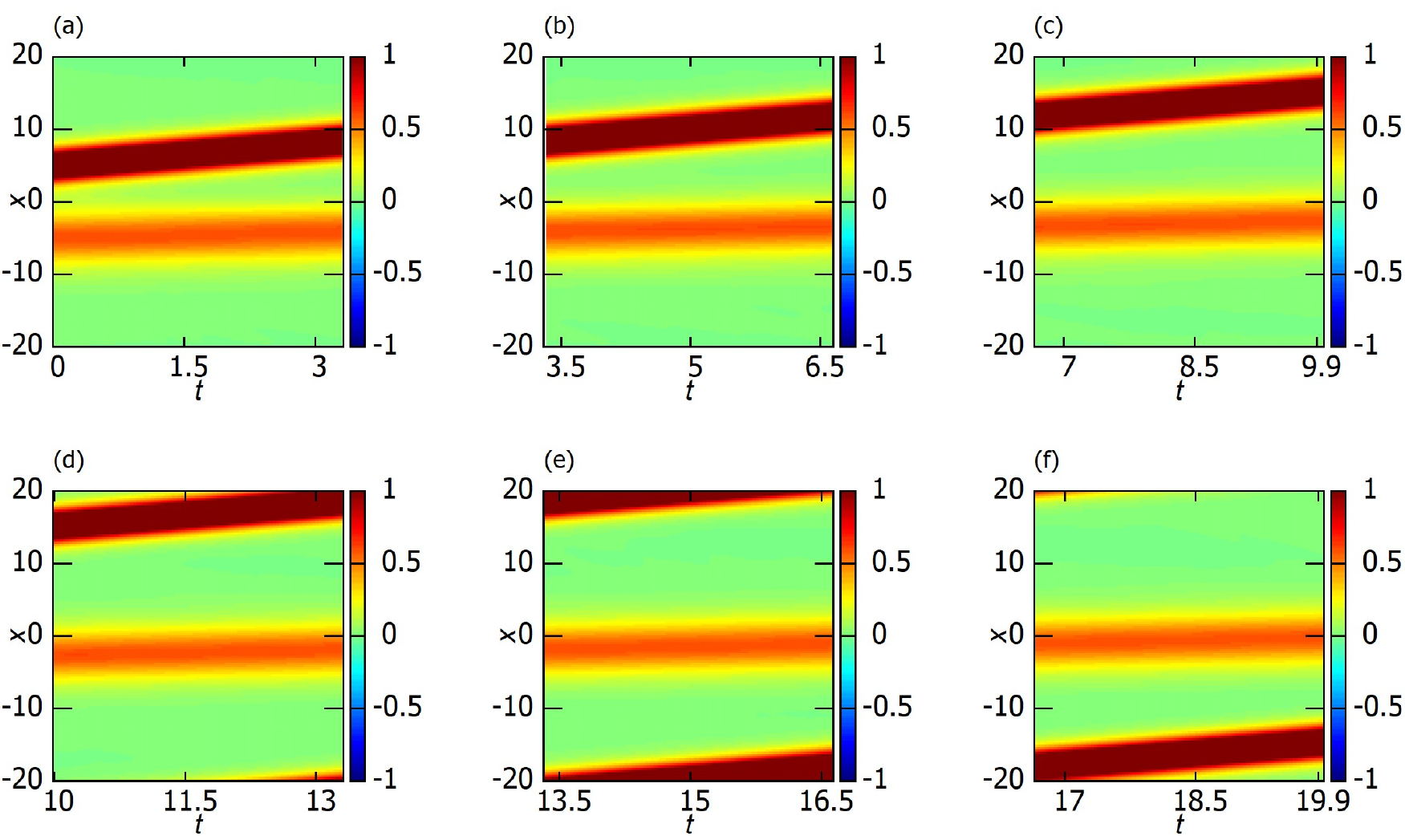}
	\caption{\label{fig2sol}2-soliton training data of $c_1=1.0,c_2=0.25,x_0=6.0$. with the time segment of 
	(a)~$t\in[0.0,3.3)$, (b)~$t\in[3.3,6.6)$, (c)~$t\in[6.6,9.9)$, (d)~$t\in[9.9,13.3)$, (e)~$t\in[13.3,16.6)$, (f)~$t\in[16.6,19.9)$.}
\end{figure}

For the next application, we consider the 2-soliton initial profile. 
This is no longer solution of the equation although each soliton is a traveling wave solution. 
We place two single solitons far enough so that they never collide with each other. 
We define the initial profile
\begin{equation}
u(x,y,0)=U_{c_1}(x-x_0,y)+U_{c_2}(x+x_0,y)\,,
\end{equation} 
where $c_1$ and $c_2$ are the speed of each soliton, respectively. 
The initial profile does not possess convertibility of the $x$ derivative and the $t$ derivative. For the training 
data of the inverse analysis, we employ the segment data: $t\in$ (a)~$[0.0,3.3)$, (b)~$[3.3,6.6)$, (c)~$[6.6,9.9)$,
(d)~$[9.9,13.3)$, (e)~$[13.3,16.6)$, (f)~$[16.6,19.9)$, as plotted in Fig.\ref{fig2sol}. The results of the inverse analysis are presented in Table \ref{tab:misidentify_ZK_2_sol}. 
The accuracy has been certainly improved, but the \textit{MSE}s still need to be improved. 
There is yet another major disadvantage in this scheme. 
It is now becoming evident that the PINN inverse analysis loses 
the resolutions when two solitons experience the collision processes, 
and this might happen in this instance~\cite{shimasaki24col}. 
In the following section, we will discuss the novel prescriptions for improving resolution that do not rely on the dynamics of the 2-soliton. 

\begin{table}[H]
	\centering
	\caption{The identification of the ZK equation in terms of the 2-soliton with the small time 
	segment data: $t\in$ (a)~$[0.0,3.3)$, (b)~$[3.3,6.6)$, (c)~$[6.6,9.9)$,(d)~$[9.9,13.3)$, (e)~$[13.3,16.6)$, (f)~$[16.6,19.9)$.}
	\label{tab:misidentify_ZK_2_sol}
	\begin{tabular}{|c|c|c|}
		\hline
		& $\mathit{PDE}$ & $\mathit{MSE}$ $(\times 10^{-6})$
		\\ \hline
		Correct equation & $u_t+2uu_x+0.0\left(\nabla^2 u\right)_t+\left(\nabla^2 u\right)_x=0$ & $-$
		\\ \hline
		Identified (a) &$u_t+2.056uu_x-0.001\left(\nabla^2 u\right)_t+1.058\left(\nabla^2 u\right)_x=0$ & 5.0
		\\ \hline
		Identified (b) &$u_t+2.046uu_x-0.029\left(\nabla^2 u\right)_t+1.020\left(\nabla^2 u\right)_x=0$ & 5.3
		\\ \hline
		Identified (c) &$u_t+2.004uu_x-0.026\left(\nabla^2 u\right)_t+1.014\left(\nabla^2 u\right)_x=0$ & 6.5
		\\ \hline
		Identified (d) &$u_t+2.032uu_x-0.039\left(\nabla^2 u\right)_t+0.997\left(\nabla^2 u\right)_x=0$ & 8.3
		\\ \hline
		Identified (e) &$u_t+1.991uu_x-0.021\left(\nabla^2 u\right)_t+0.996\left(\nabla^2 u\right)_x=0$ & 9.8
		\\ \hline
		Identified (f) &$u_t+2.050uu_x-0.005\left(\nabla^2 u\right)_t+1.060\left(\nabla^2 u\right)_x=0$ & 5.5
		\\ \hline
	\end{tabular}
\end{table}

\section{\label{sec:4}Data-driven discovery of the governing equations}

In this section, we would like to examine the equations predicted from the data of the single vortex with some non-trivial ``twists''.
It is unquestionably useful because observational data increase various noises, modulations to some degree, and other disturbances to some degree. 
Therefore, the attempt to derive the governing equation using PINNs from data with contamination. 
We give a few examples that successfully identify the governing equations.

\subsection{Deformation of the initial profile}

\begin{figure}[t]
	\centering
		\includegraphics[width=1.0\linewidth]{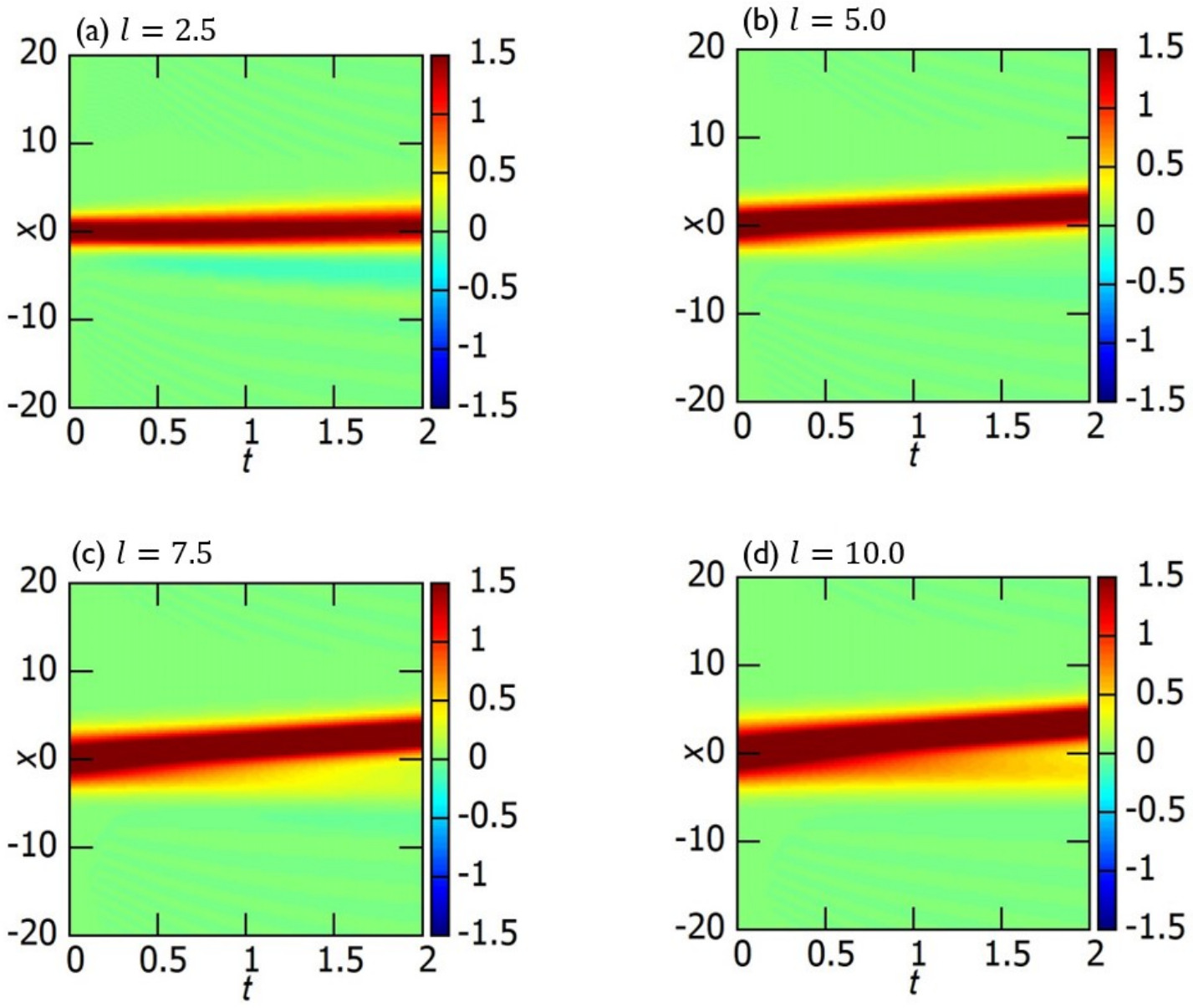}\\
	\caption{\label{fig:gaussian_ZK} The time evolution of the Gaussian profiles with $\ell=2.5,5.0,7.5$, and $10.0$.}
\end{figure}

One of the reasons for the misidentification is that we employed the traveling-wave solution of the normalized 
equation~\eqref{normalizedeq} as the initial profile.
A non-linear PDE often has numerous independent solutions caused by differences in the initial 
conditions~\cite{CHEN2024102041}, providing deeper insight into the mathematical meanings of 
the equation and the phenomena of our concern.  
If we use the data with the various initial profiles in the PINN inverse analysis, 
the resolution improves because
this prescription would break the condition $\frac{\partial u}{\partial x}=-c\,\frac{\partial u}{\partial t}$. 
The Gaussian profile
\begin{equation}
u(x,y,0):=2.0\exp\left[-\frac{1}{\ell}(x^2+y^2)\right]
\label{gaussian}
\end{equation}
is widely used in the analysis of several evolution equations in two spatial dimensions.  
It also can be an initial profile of the ZK and the RLW equations, so we numerically solve these governing equations and get the data for the inverse analysis. 
We present the results the for forward analysis of the ZK equation with the initial profiles~(\ref{gaussian}) of the several scales $\ell=2.5,5.0,7.5$ and $10.0$ in Figure \ref{fig:gaussian_ZK}. 
The results of the inverse analysis for the several data with $\ell$ from 2.25 to 10.0 are given in Table \ref{tab:identify_gaussian}. 
The results show that, as the \textit{MSE} is not small enough ($\sim 10^{-5}$), further convergence has to be attained.
  
\begin{table}[H]
	\centering
	\caption{The identification of the ZK equation using the data of the Gaussian initial profiles.}
	\label{tab:identify_gaussian}
	\begin{tabular}{|c|c|c|}
		\hline
		& $\mathit{PDE}$ & $\mathit{MSE}$ $(\times 10^{-5})$
		\\ \hline
		Correct equation & $u_t+2uu_x+0.0\left(\nabla^2 u\right)_t+\left(\nabla^2 u\right)_x=0$ & $-$
		\\ \hline
		Identified $\ell=2.25$ &$u_t+1.986uu_x-0.0581\left(\nabla^2 u\right)_t+0.9797\left(\nabla^2 u\right)_x=0$ & $6.1$
		\\ \hline
		Identified $\ell=2.50$ &$u_t+2.0407uu_x-0.0569\left(\nabla^2 u\right)_t+1.0106\left(\nabla^2 u\right)_x=0$ & $3.7$
		\\ \hline
		Identified $\ell=2.75$ &$u_t+2.0558uu_x-0.0771\left(\nabla^2 u\right)_t+1.0102\left(\nabla^2 u\right)_x=0$ & $2.3$
		\\ \hline
		Identified $\ell=4.75$ &$u_t+2.0624uu_x-0.1149\left(\nabla^2 u\right)_t+0.94666\left(\nabla^2 u\right)_x=0$ & $3.8$
		\\ \hline
		Identified $\ell=5.0$ &$u_t+2.0599uu_x-0.14533\left(\nabla^2 u\right)_t+0.92023\left(\nabla^2 u\right)_x=0$ & $3.0$
		\\ \hline
		Identified $\ell=5.25$ &$u_t+2.05069uu_x-0.06841\left(\nabla^2 u\right)_t+0.97961\left(\nabla^2 u\right)_x=0$ & $2.4$
		\\ \hline
		Identified $\ell=7.25$ &$u_t+2.0534uu_x-0.0475\left(\nabla^2 u\right)_t+0.9973\left(\nabla^2 u\right)_x=0$ & $1.8$
		\\ \hline
		Identified $\ell=7.50$ &$u_t+2.05035uu_x-0.04637\left(\nabla^2 u\right)_t+0.9950\left(\nabla^2 u\right)_x=0$ & $1.7$
		\\ \hline
		Identified $\ell=7.75$ &$u_t+2.04880uu_x-0.052728\left(\nabla^2 u\right)_t+0.9840\left(\nabla^2 u\right)_x=0$ & $2.0$
		\\ \hline
		Identified $\ell=10.0$ &$u_t+2.0473uu_x-0.0369\left(\nabla^2 u\right)_t+0.9976\left(\nabla^2 u\right)_x=0$ & $1.3$
		\\ \hline
	\end{tabular}
\end{table}

\begin{figure}[t]
	\centering
	\includegraphics[width=1.0\linewidth]{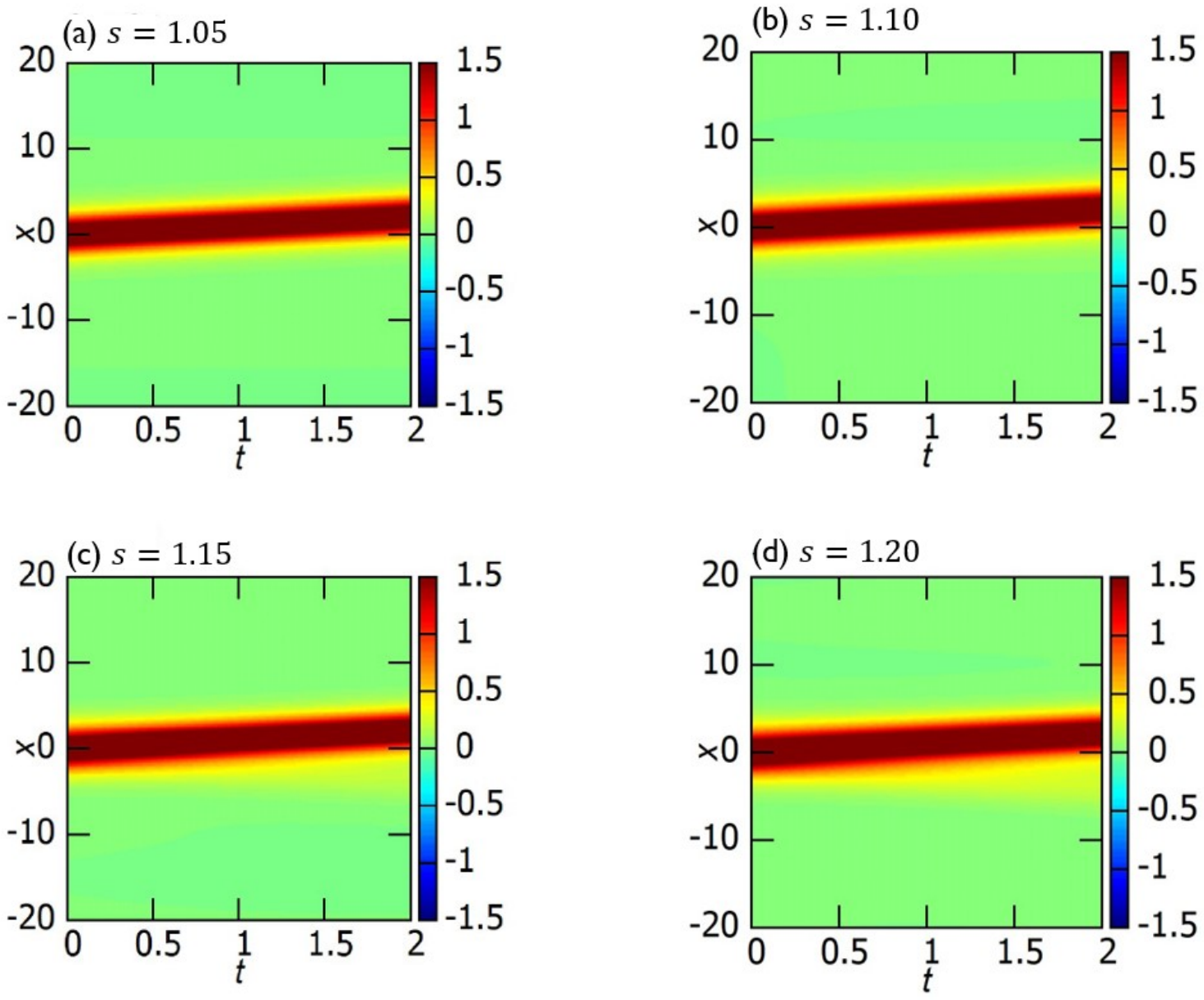}\\
	\caption{\label{fig:oval_ZK} The time evolution of the oval-shaped profiles with $s=1.05,1.10,1.15$, and $1.20$.}
\end{figure}

Another initial profile might have more physical meaning. 
In geophysical flow dynamics as seen in Jupiter's GRS, the oval-shaped profile is frequently observed. 
Also, the ZK equation is a toy model of the intermediate geostrophic regime, which is believed to govern the GRS.
We employ the following form of the initial condition 
\begin{equation}
u(x,y,0):=U\biggl(s\,\xi,~~\frac{1}{s}y\biggr)\,,~~s> 1\,.
\label{oval}
\end{equation}
Again it breaks the coincidence $\frac{\partial u}{\partial x}=-c\,\frac{\partial u}{\partial t}$, 
then improves the PINNs' resolution. We present the results for the forward analysis of the ZK equation with the initial profile~(\ref{oval}) with the several deformation parameter values $s=1.05,1.1,1.15$, and $1.20$ in Fig. \ref{fig:oval_ZK}. 
The results of the inverse analysis for the several data with $s$ from 1.0 to $1.2$ are 
in Table \ref{tab:identify_oval}.
Apparently, the predictive power of the PINNs is certainly improved for $s>1.125$. 
At the same time, we must admit that the current prescription has a downside that excessive deformation often contaminates the vortex later. 
The origin of the instability is that it inevitably violates the conservation laws.  
We conclude that while using deformed initial profiles is somewhat promising, but 
it is only if the deviation from the circular shape of the initial profile is small. 

\begin{figure}[htbp]
\includegraphics[width=0.4\linewidth]{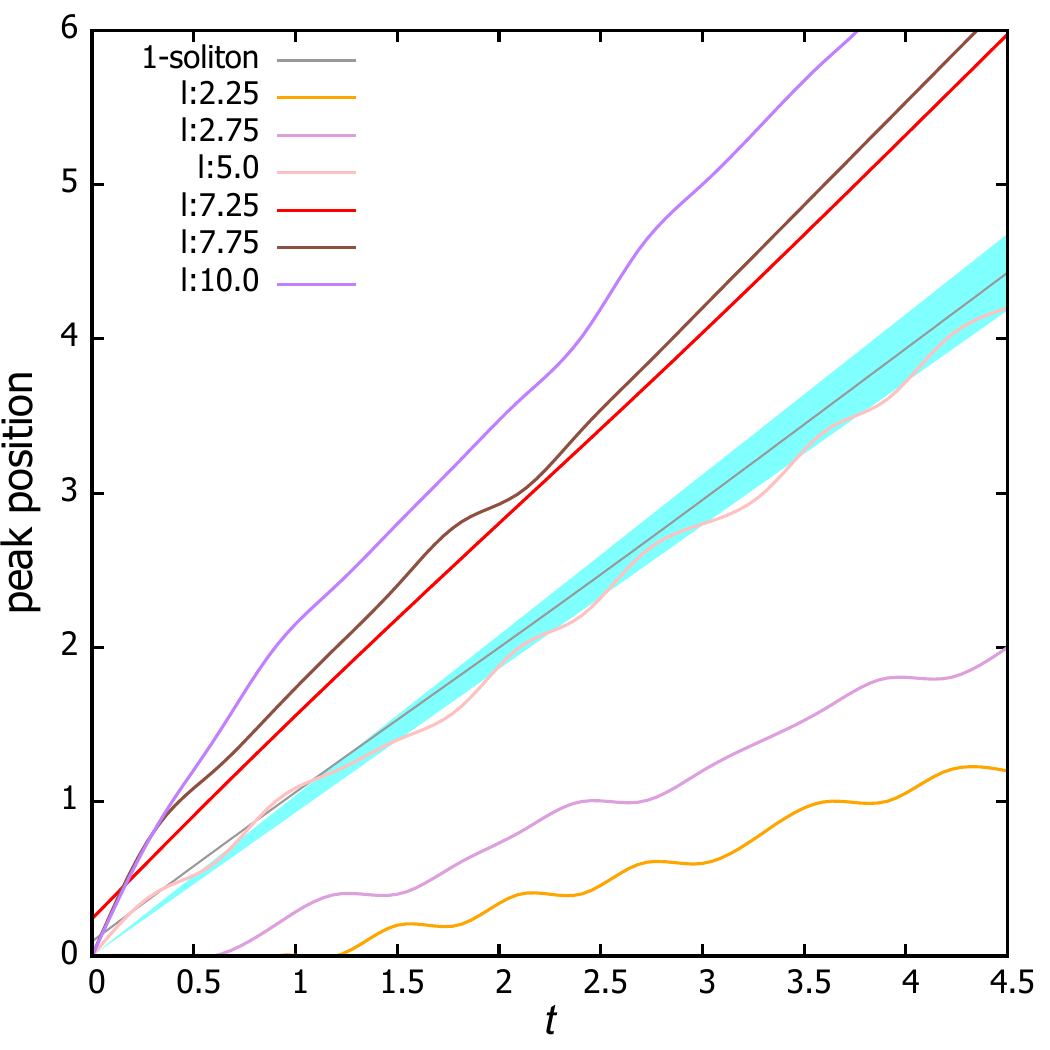}~~~~
\includegraphics[width=0.4\linewidth]{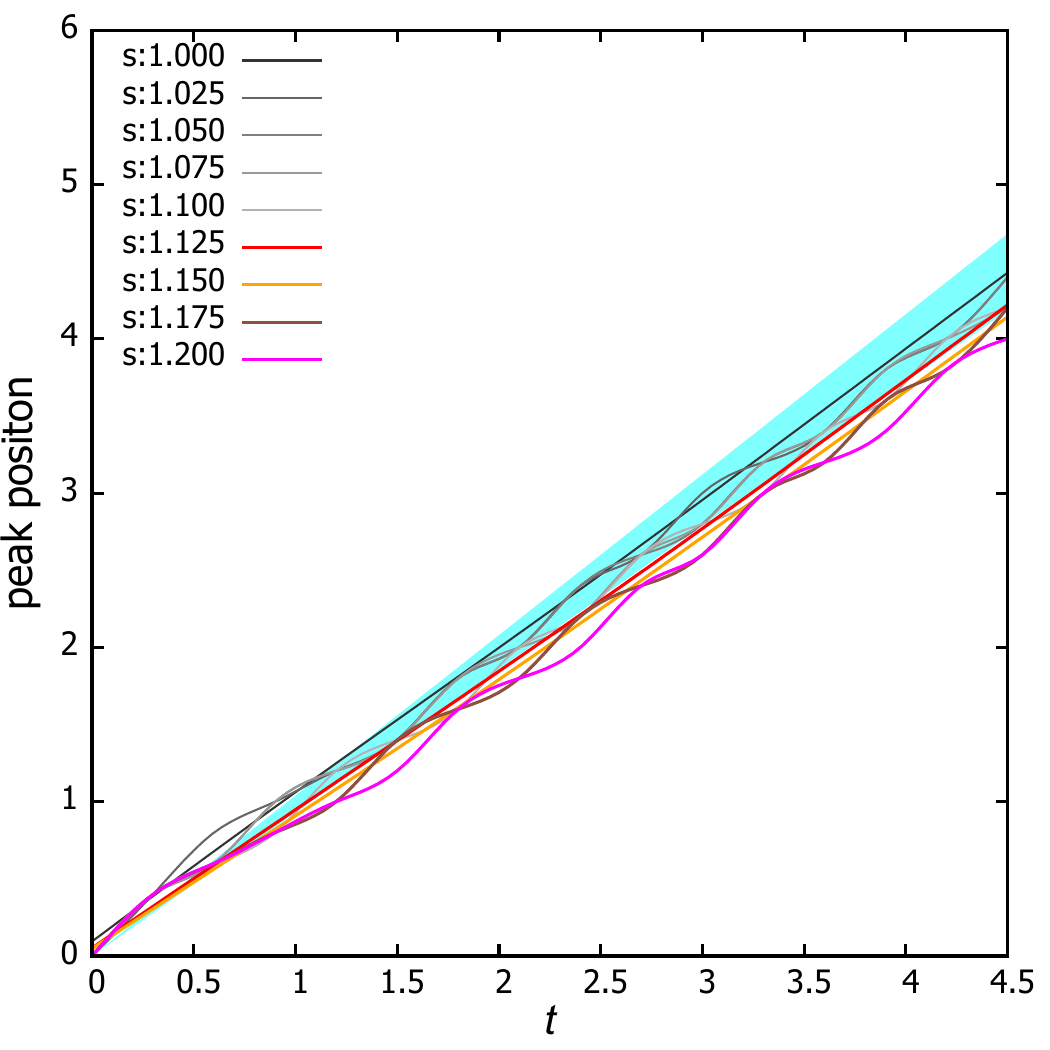}
\caption{\label{velocity_deformation}The position of the center of the vortices of 
the Gaussian deformation (left) and the oval-shaped deformation (right). The blue-shaded
areas indicate that the identification of the equations is unsuccessful,  roughly the relation $u_t\sim-u_x$ is satisfied.
}
\end{figure}

Here, it is worthwhile to address why the resolution of the PINNs has been improved. 
In Fig.\ref{velocity_deformation}, we plot the velocity of the center of the vortices in the case of Gaussian and oval profiles.  
The blue-shaded areas indicate where the identifications of the equations fail. 
The speed of the vortices appears to be crucial; the identification is only successful when the velocities are outside the blue regions.

\begin{table}[H]
	\centering
	\caption{The identification of the ZK equation with the data of the oval-shaped initial profiles.}
	\label{tab:identify_oval}
	\begin{tabular}{|c|c|c|}
		\hline
		& $\mathit{PDE}$ & $\mathit{MSE}$ $(\times 10^{-6})$
		\\ \hline
		Correct equation & $u_t+2uu_x+0.0\left(\nabla^2 u\right)_t+\left(\nabla^2 u\right)_x=0$ & $-$
		\\ \hline
		Identified $s=1.000$ &$u_t+1.995uu_x-0.0581\left(\nabla^2 u\right)_t+0.45367\left(\nabla^2 u\right)_x=0$ & $1.5$
		\\ \hline
		Identified $s=1.025$ &$u_t+2.0140uu_x-0.5489\left(\nabla^2 u\right)_t+0.45300\left(\nabla^2 u\right)_x=0$ & $3.4$
		\\ \hline
		Identified $s=1.050$ &$u_t+2.0241uu_x-0.4677\left(\nabla^2 u\right)_t+0.53675\left(\nabla^2 u\right)_x=0$ & $6.8$
		\\ \hline
		Identified $s=1.075$ &$u_t+2.02010uu_x-0.4754\left(\nabla^2 u\right)_t+0.51766\left(\nabla^2 u\right)_x=0$ & $11.8$
		\\ \hline
		Identified $s=1.100$ &$u_t+2.0248uu_x-0.3649\left(\nabla^2 u\right)_t+0.64731\left(\nabla^2 u\right)_x=0$ & $13.6$
		\\ \hline
		Identified $s=1.125$ &$u_t+1.99602uu_x-0.07228\left(\nabla^2 u\right)_t+0.92303\left(\nabla^2 u\right)_x=0$ & $9.5$
		\\ \hline
		Identified $s=1.150$ &$u_t+2.00005uu_x-0.08176\left(\nabla^2 u\right)_t+0.91783\left(\nabla^2 u\right)_x=0$ & $10.2$
		\\ \hline
		Identified $s=1.175$ &$u_t+1.9961uu_x-0.06765\left(\nabla^2 u\right)_t+0.92731\left(\nabla^2 u\right)_x=0$ & $13.0$
		\\ \hline
		Identified $s=1.200$ &$u_t+1.9975uu_x-0.03614\left(\nabla^2 u\right)_t+0.9611\left(\nabla^2 u\right)_x=0$ & $12.8$
		\\ \hline
	\end{tabular}
\end{table}

\subsection{A perturbative friction via a background flow}
Both the RLW and the ZK equations are governing equations of the plasma or the fluid dynamics. 
In fact, the former is a simplified form of the drift wave equation in plasma or the quasi-geostrophic equation describing Rossby waves in rotating fluid when we omit the Jacobian-type non-linear term from the governing equation. 
The latter is closely connected with the intermediate geostrophic regime of the shallow water dynamics~\cite{Williams84}.
The equation possesses the anti-cyclonic stable vortex, which is supposed to be a candidate of the GRS. 
For both the drift wave equation and the intermediate geostrophic equation, the existence of the shear flows is crucial for obtaining the stable, long-lived solutions~\cite{SUTYRIN2021101782,Williams84,Koike:2022gfq}. 

\begin{figure}[t]
	\centering
	\includegraphics[width=1.0\linewidth]{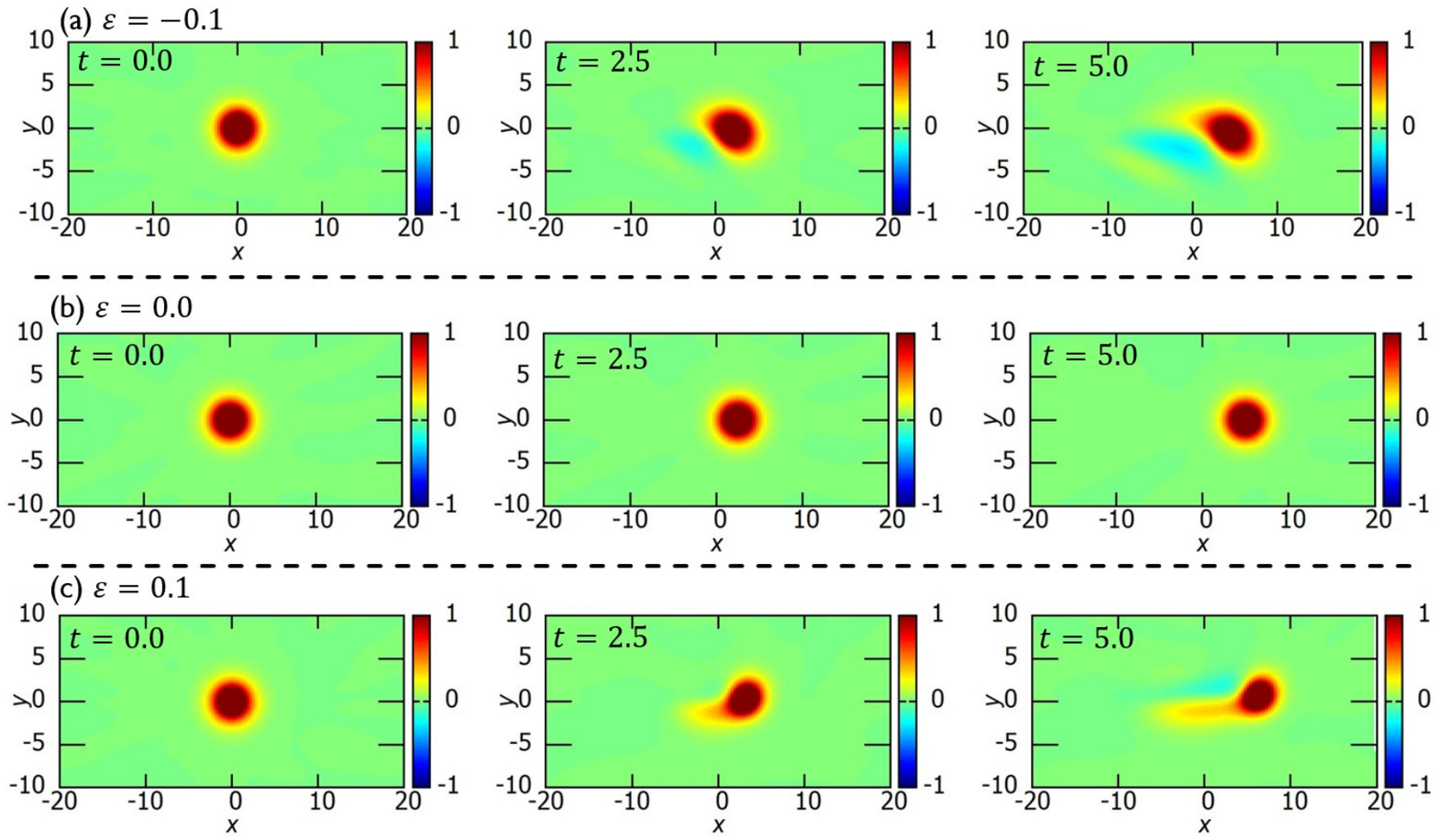}
	\caption{\label{fig:ep_const}The training data with constant current $u_0=-1.0+\varepsilon$, 
	where (a): $\varepsilon=-0.1$,(b): $\varepsilon=0.0$,(c): $\varepsilon=0.1$.}
\end{figure}

\begin{figure}[t]
	\centering
	\includegraphics[width=1.0\linewidth]{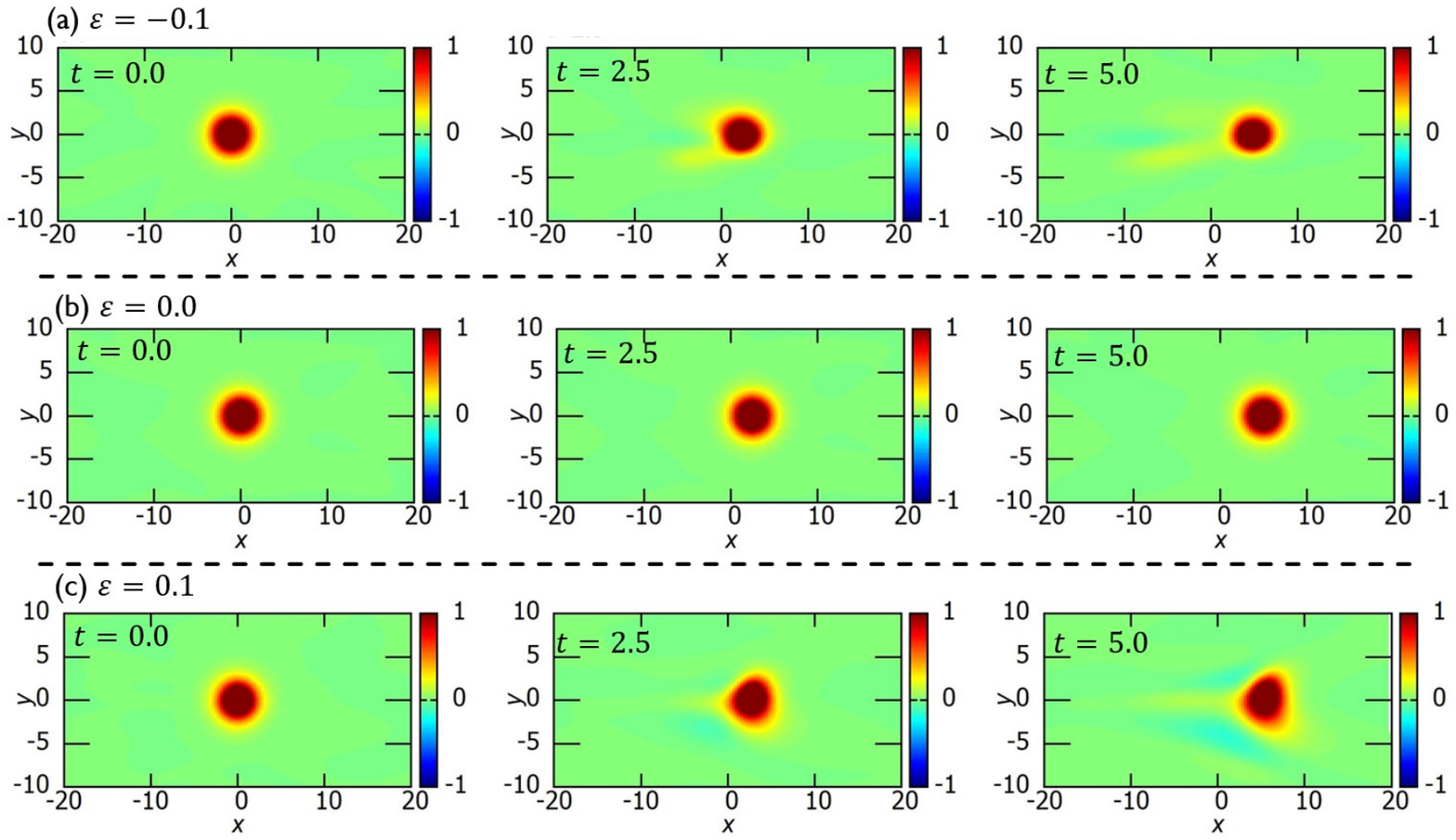}
	\caption{\label{fig:ep_y}The training data with $y$-dependent flow $u_0=-1.0+\varepsilon y$, 
	where (a): $\varepsilon=-0.1$,(b): $\varepsilon=0.0$,(c): $\varepsilon=0.1$.}
\end{figure}

Our goal here is to enhance the PINNs' performance by utilizing the influence of the background flow.  
For this purpose, we use the above intermediate geostrophic equation to find the perturbative friction term. 
The equation for the cyclonic shear~\cite{Williams84,charney1981oceanic} is 
\begin{align}
\frac{\partial \eta}{\partial t}+2\eta\frac{\partial \eta}{\partial x}-\frac{\partial}{\partial x}(\nabla^2\eta)
+2y\frac{\partial \eta}{\partial x}-2J[\nabla^2\eta,\eta]=0\,.
\label{WYF}
\end{align}
The last term, the Jacobian represents the geostrophic advection of vorticity \(\nabla^2\eta\) 
in the words in fluid mechanics, and we refer to it as the vector nonlinear term.
The stream function $\eta:=\eta(x,y,t)$ is expanded by the shear zonal flow $u_0(y)$ such as
\begin{align}
\eta(x,y,t)=\int_0^y u_0(y')dy'+u(x,y,t)\,,
\label{eta}
\end{align}
where the field \(u(x,y,t)\) is a perturbation from the sole existence of the zonal flow. 
If we set $u_0=-1.0$ and omit the Jacobian, the equation becomes the ZK equation~\cite{Koike:2022gfq}. 
Setting the constant current \(u_0\) in (\eqref{eta}) as 
\begin{align}
u^1_0:=-1.0+\varepsilon,~~|\varepsilon|\ll 1\,,
\end{align}
we can write down \eqref{WYF} in terms of $u$ such as
\begin{align}
u_t+2uu_x+(\nabla^2u)_x-2J[\nabla^2u,u]+2\varepsilon \{yu_x-(\nabla^2u)_x)\}=0\,.
\end{align}
After the removal of the Jacobian, the perturbative term which we use later becomes 
\begin{equation}
\varepsilon\,\mathcal{N}^1(u_x,u_{xxx},u_{xyy}):=2\varepsilon \{yu_x-\left(\nabla^2u\right)_x\}.
\label{perturb0}
\end{equation}
Similarly, if we consider a $y$-dependent shear flow and set \(u^2_0(y)\) as 
\begin{align}
u^2_0(y):=-1.0+\varepsilon y,~~|\varepsilon|\ll 1,
\end{align}
we find
\begin{align}
\varepsilon\,\mathcal{N}^2(u_x,u_{xxx},u_{xyy}):=\varepsilon \{y^2u_x-2y\left(\nabla^2u\right)_x\}.
\label{perturb1}
\end{align}
The modified ZK equation is defined as
\begin{align}
\mathcal{F}^a_\textrm{mZK}=u_t+\mathcal{N}_\textrm{ZK}(u,u_x,u_{xxx},u_{xyy})+\varepsilon\mathcal{N}^a(u_x,u_{xxx},u_{xyy})=0,~~a=1,2\,.
\label{mZK}
\end{align} 
Note that the equation of $\mathcal{F}^1_\textrm{mZK}$ 
possesses four conserved quantities, where the first two $I_1,I_2$ is completely same as those of 
the ZK equation while the other conserved quantities $I_3,\bm{I}_4$ are slightly 
modified 
from \eqref{ZKCQ3} and \eqref{ZKCQ4} to
\begin{align}
&I_3':=\int \biggl[(\nabla u)^2-\frac{1}{1-2\varepsilon}\frac{1}{3}u^3-\frac{\varepsilon}{1-2\varepsilon}yu^2\biggr]dxdy\,,
\label{ZKCQ3e}
\\
&\bm{I}_4':=\int\bm{r}udxdy-t\bm{e}_x\int (u^2+2\varepsilon yu)dxdy\,.
\label{ZKCQ4e}
\end{align}
The $I_3'$ corresponds to the total energy in $\mathcal{F}^1_\textrm{mZK}$.
It is simply obtained on multiplying $\nabla^2u$ by the equation and using the doubly periodic boundary condition to integrate it on the whole area. 
It comprises the kinetic energy:~$(\nabla u)^2$, a potential energy:~$u^3$, and also a twisting term:~$yu^2$. 
Note that the $|\varepsilon|$ is small, so the denominators in (\ref{ZKCQ3e}) never become zero. 
The solutions of the equation $\mathcal{F}^1_\textrm{mZK}$ are stable for the time evolution 
because the energy is conserved. 
On the other hand, the energy of the $\mathcal{F}^2_\textrm{mZK}$ is not conserved, and the solutions tend to be unstable.
We present the results of the forward analysis of $\mathcal{F}^1_\textrm{mZK}$ in Fig.\ref{fig:ep_const}
and $\mathcal{F}^2_\textrm{mZK}$ in Fig.\ref{fig:ep_y}. 
Here, we plot the solutions with the strength of the perturbative terms
$\varepsilon = -1.0, 0.0$, and $1.0$. 

\begin{figure}[htbp]
	\begin{center}
		\includegraphics[width=0.7\linewidth]{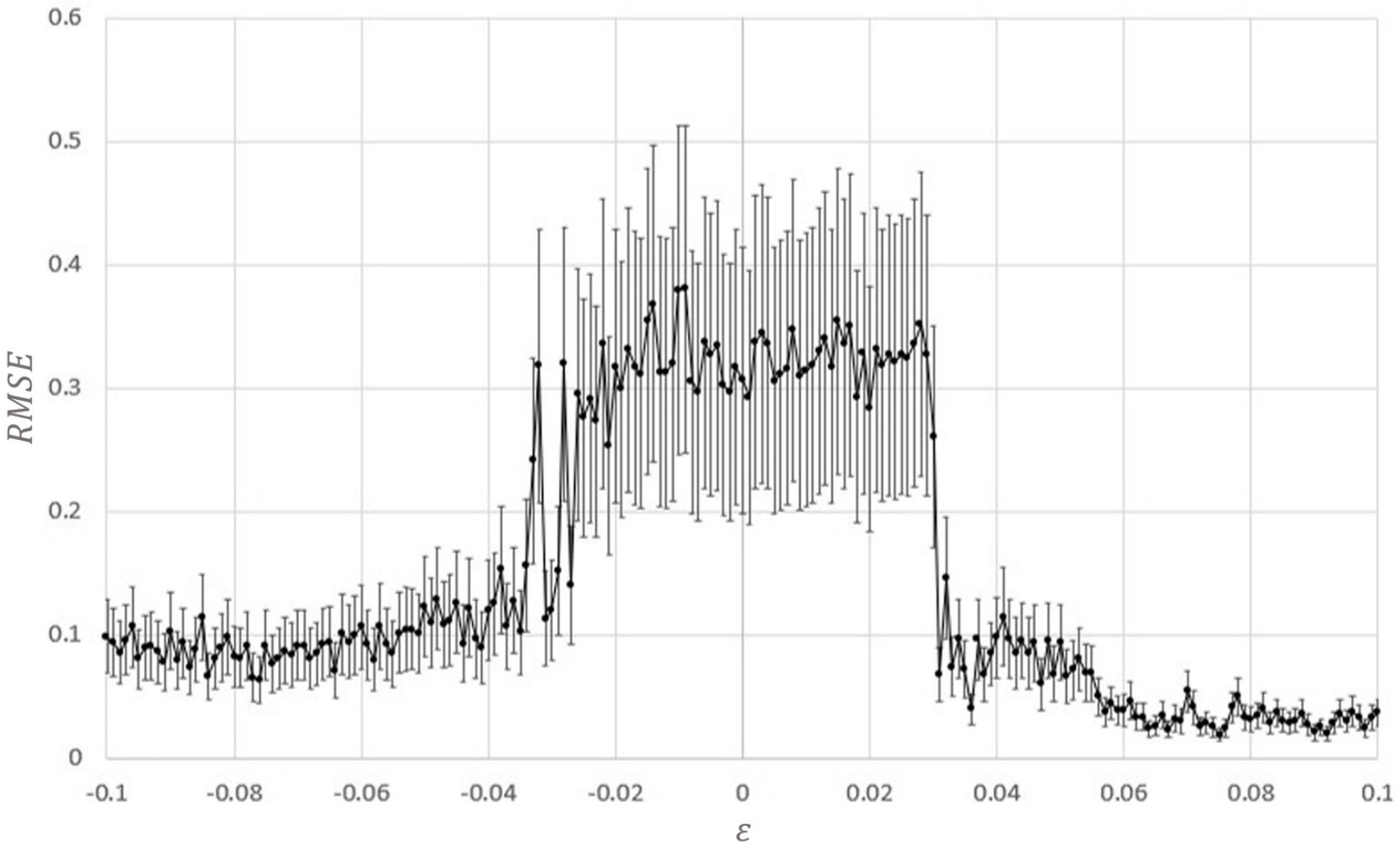}\\
		\textrm{(a)~$u_0:=-1.0+\varepsilon$}\\
		\vspace{1.5cm}
		
		\includegraphics[width=0.7\linewidth]{FIG_fric_zk_2}\\
		\textrm{(b)~$u_0:=-1.0+\varepsilon y$}
	\end{center}
	\caption{\label{Perturbation}Relation between the accuracy of the identification and strength of the perturbative term induced 
		to the ZK equation $\varepsilon$. 
		The root mean-squared error: $\textit{RMSE}=\sqrt{\dfrac{\sum_{i=0}^2|\lambda_i^\textrm{pred}-\lambda_i^\textrm{correct}|^2}{3}}$ is: 
		(a)~suppressed for $\epsilon \gtrsim 0.05$ in case of the constant current,  
		(b)~ineffective suppression of the \textit{RMSE} in the case of the $y$-dependent current. In this calculation, the inverse problem analysis is performed five times for the configurations obtained at each $\varepsilon$. The average of the RMSEs is drawn by the solid black line, and the standard uncertainty for each is represented by the gray line.}
\end{figure}

The inverse analysis is operated with
\begin{align}
&\tilde{\mathcal{F}}=u_t+\bar{\mathcal{N}}(u,u_x,u_{txx},u_{tyy},u_{xxx},u_{xyy},\bm{\lambda})+\epsilon\mathcal{N}_1(u_x,u_{xxx},u_{xyy})=0\,,
\\
&\bar{\mathcal{N}}(u,u_x,u_{txx},u_{tyy},u_{xxx},u_{xyy},\lambda_0,\lambda_1,\lambda_2):=\lambda_0uu_x+\lambda_1(\nabla^2u)_t+\lambda_2(\nabla^2u)_x\,.
\end{align}
The results are shown in Fig.\ref{Perturbation}. 
Using the results of the forward analysis to obtain the solution of \eqref{mZK}, 
we compute the root mean-squared error of the model parameters defined as 
\begin{align}
\textit{RMSE}:=\sqrt{\dfrac{\sum_{i=0}^2|\lambda_i^\textrm{pred}-\lambda_i^\textrm{correct}|^2}{3}}\,.
\end{align} 
We find that significant improvement in the resolution of the predictability is realized in the case considering the constant current with the certain perturbation. 
However, case of the $y$-dependent current, which changes the original ZK equation into a non-conservative equation, cannot improve the predictability of the PINNs.
As a result, we conclude that PINN predictability can be increased with a small perturbation that keeps the conservation laws.

\clearpage

\begin{table}[H]
	\centering
	\caption{\label{zk_id_fric_cons_0} The identification of the ZK equation using the vortex data with constant current $\varepsilon=0.001$.}
	\centering
	\begin{tabular}{|c|c|}
		\hline
		& $\mathit{PDE}$\\ \hline
		\textrm{Correct equation}~& $u_t+2uu_x+\left(\nabla^2 u\right)_x=0$ \\ \hline
		\textrm{Identified(1)} & $u_t+1.996uu_x+0.5476\left(\nabla^2 u\right)_x-0.4503\left(\nabla^2 u\right)_t=0$\\ \hline
		\textrm{Identified(2)} & $u_t+1.993uu_x+0.7105\left(\nabla^2 u\right)_x-0.2839\left(\nabla^2 u\right)_t=0$\\ \hline
		\textrm{Identified(3)} & $u_t+1.997uu_x+0.5657\left(\nabla^2 u\right)_x-0.4326\left(\nabla^2 u\right)_t=0$\\ \hline
		\textrm{Identified(4)} & $u_t+1.999uu_x+0.3839\left(\nabla^2 u\right)_x-0.6170\left(\nabla^2 u\right)_t=0$\\ \hline
		\textrm{Identified(5)} & $u_t+1.996uu_x+0.5573\left(\nabla^2 u\right)_x-0.4433\left(\nabla^2 u\right)_t=0$\\ \hline
		\textrm{Standard uncertainty of \textit{RMSE}}& 0.108\\ \hline
	\end{tabular}
\end{table} 

\begin{table}[H]
	\centering
	\caption{\label{zk_id_fric_cons_1} The identification of the ZK equation using the vortex data with constant current $\varepsilon=0.04$.}
	\centering
	\begin{tabular}{|c|c|}
		\hline
		& $\mathit{PDE}$\\ \hline
		\textrm{Correct equation}~& $u_t+2uu_x+\left(\nabla^2 u\right)_x=0$ \\ \hline
		\textrm{Identified(1)} & $u_t+2.016uu_x+1.136\left(\nabla^2 u\right)_x+0.1535\left(\nabla^2 u\right)_t=0$\\ \hline
		\textrm{Identified(2)} & $u_t+2.023uu_x+1.143\left(\nabla^2 u\right)_x+0.1646\left(\nabla^2 u\right)_t=0$\\ \hline
		\textrm{Identified(3)} & $u_t+2.009uu_x+1.132\left(\nabla^2 u\right)_x+0.1452\left(\nabla^2 u\right)_t=0$\\ \hline
		\textrm{Identified(4)} & $u_t+2.001uu_x+1.124\left(\nabla^2 u\right)_x+0.1387\left(\nabla^2 u\right)_t=0$\\ \hline
		\textrm{Identified(5)} & $u_t+2.019uu_x+1.125\left(\nabla^2 u\right)_x+0.1233\left(\nabla^2 u\right)_t=0$\\ \hline
		\textrm{Standard uncertainty of \textit{RMSE}}& \bm{0.031}\\ \hline
	\end{tabular}
\end{table} 

\begin{table}[H]
	\centering
	\caption{\label{zk_id_fric_cons_2} The identification of the ZK equation using the vortex data with constant current $\varepsilon=0.1$.}
	\centering
	\begin{tabular}{|c|c|}
		\hline
		& $\mathit{PDE}$\\ \hline
		\textrm{Correct equation}~& $u_t+2uu_x+\left(\nabla^2 u\right)_x=0$ \\ \hline
		\textrm{Identified(1)} & $u_t+2.014uu_x+1.040\left(\nabla^2 u\right)_x+0.035\left(\nabla^2 u\right)_t=0$\\ \hline
		\textrm{Identified(2)} & $u_t+2.091uu_x+1.011\left(\nabla^2 u\right)_x+0.026\left(\nabla^2 u\right)_t=0$\\ \hline
		\textrm{Identified(3)} & $u_t+2.013uu_x+1.013\left(\nabla^2 u\right)_x+0.018\left(\nabla^2 u\right)_t=0$\\ \hline
		\textrm{Identified(4)} & $u_t+2.002uu_x+1.021\left(\nabla^2 u\right)_x+0.027\left(\nabla^2 u\right)_t=0$\\ \hline
		\textrm{Identified(5)} & $u_t+2.012uu_x+1.025\left(\nabla^2 u\right)_x+0.022\left(\nabla^2 u\right)_t=0$\\ \hline
		\textrm{Standard uncertainty of \textit{RMSE}}& \bm{0.009}\\ \hline
	\end{tabular}
\end{table} 

\clearpage

\begin{figure}[htbp]
\includegraphics[width=0.4\linewidth]{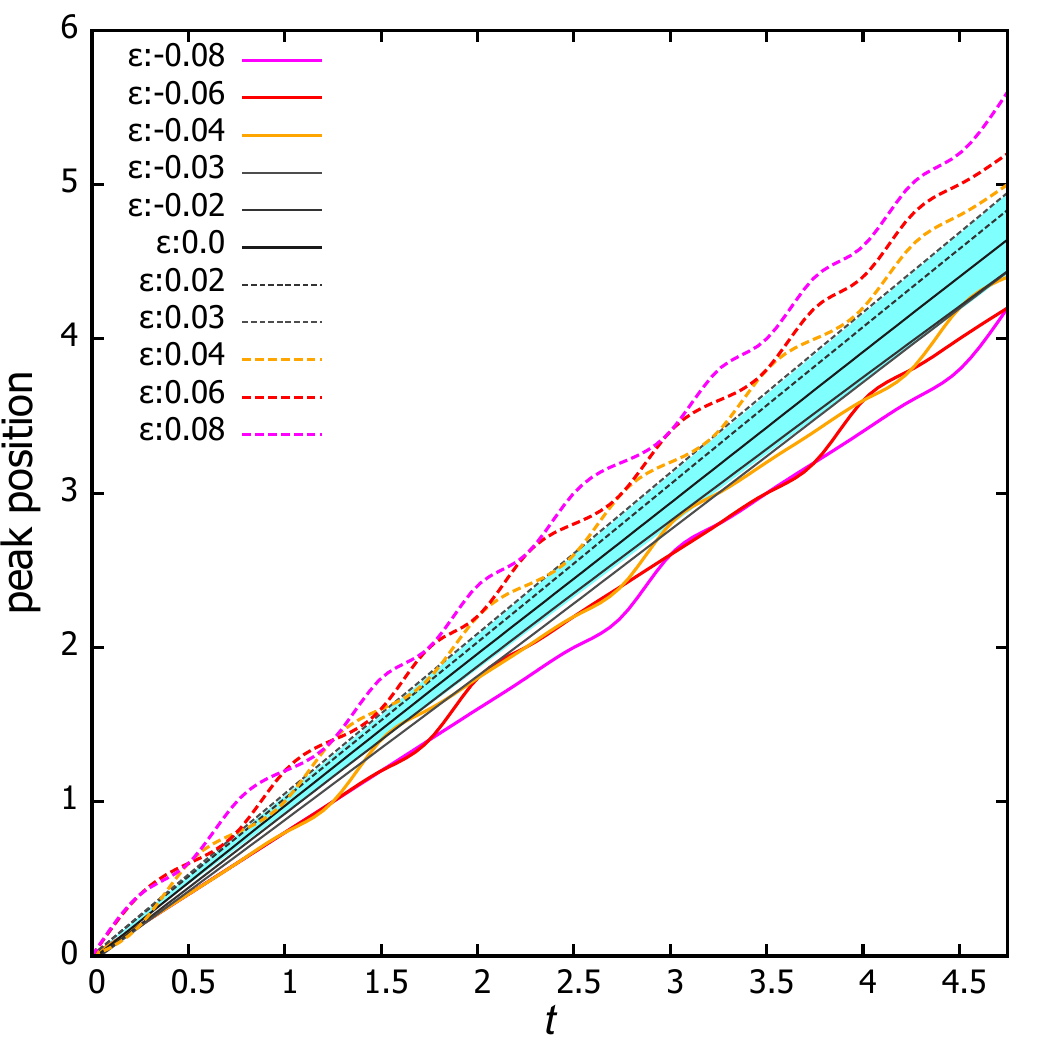}~~~~
\includegraphics[width=0.4\linewidth]{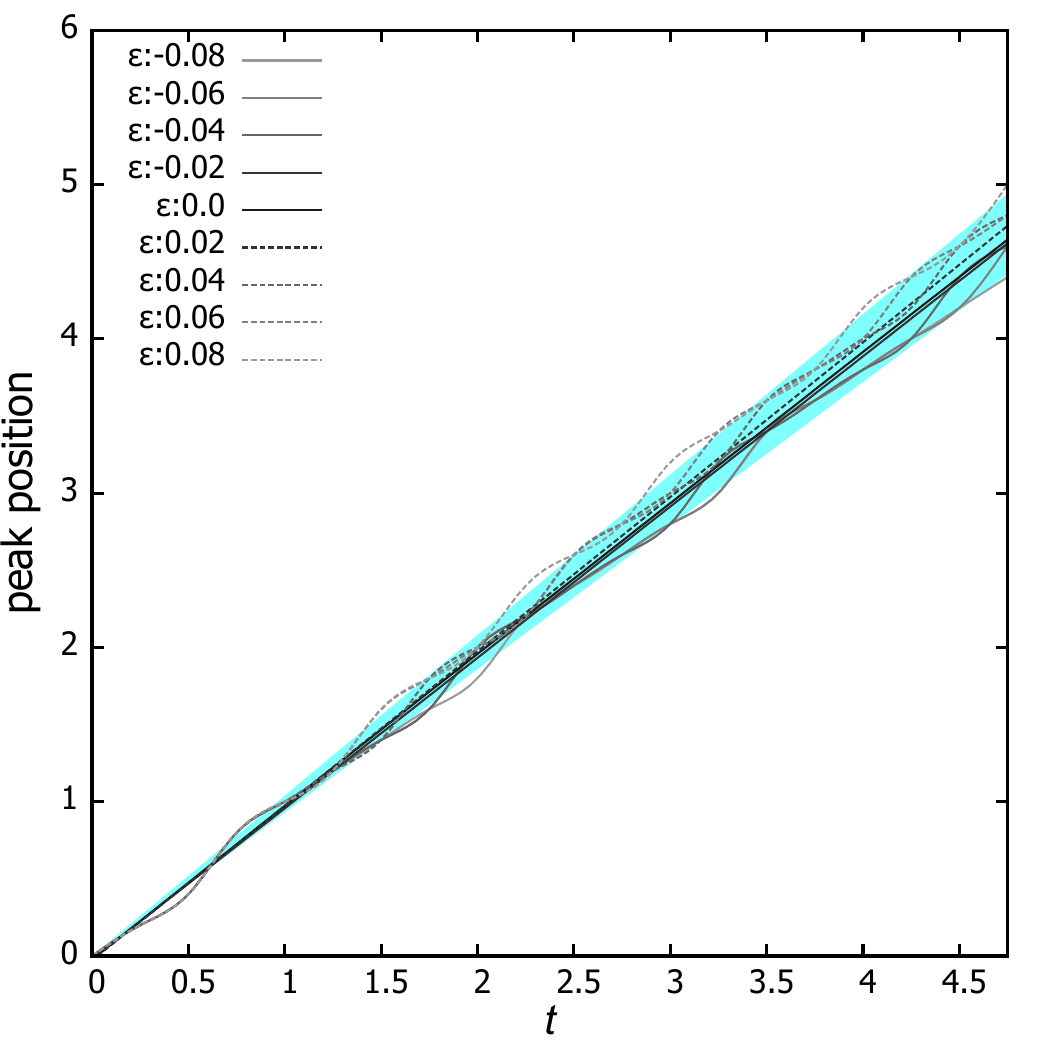}
\caption{\label{velocity_pert}
The position of the center of the vortices of 
the constant current (left) and the $y$-dependent current (right). The blue-shaded
areas indicate that the identification of the equations is unsuccessful, roughly 
the relation $u_t\sim-u_x$ is satisfied.}
\end{figure}

To quantitatively evaluate the improvement in resolution of the PINNs, we performed the inverse problem analysis five times using the vortex data corresponding to three different values of $\varepsilon$ to compute \textit{RMSE}. 
The standard error of the \textit{RMSE} represents the degree of misidentification: In FIG.\ref{Perturbation}, the error-bars indicate the uncertainty in the equation determination.
These results suggest that the vortex data considering constant current with $|\varepsilon|\gtrsim 0.04$ produce several times smaller errors in the inverse problem than the other data.
Fig.\ref{velocity_pert} again shows the velocities of the center of the vortices 
with the perturbations. 
One can directly see that in the case of the \(y\)-dependent current, the solutions always stick into the blue-shaded area, and then, the resolution fails to recover. 
Finally, we show more explicitly the behavior depending on the coefficients of the constant current with some values of $\varepsilon$. 
In Tables \ref{zk_id_fric_cons_0} - \ref{zk_id_fric_cons_2}, 
we present the results of the identification for the ZK equation and the uncertainty of the identification.

\section{\label{sec:5}Summary}

In this paper, we investigated the PINNs to analyze the quasi-integrable non-linear differential equations: 
the Zakharov-Kuznetsov equation and the regularized long-wave equation. These equations share some common structures 
for their traveling-wave solutions, and then the PINNs fail to identify them from the data. 
The identifications are improved by using the various beginning profiles, 
such as the Gaussian and the two-solitons, though the \textit{MSE}s still need to be fully converged. 
The cPINNs, where the conservation laws are applied to the \textit{MSE}, did not work in this analysis because the strength of the conservation laws appears to be too low to enhance convergence. 
If the method eventually proves effective, we will undertake a more thorough investigation 
with substantial computational resources.  

On the other hand, if we introduce a small perturbation, the friction term, into the equation, reasonable numerical convergence is observed well, and the identification is significantly improved by the other approaches.  
`Primarily, it works well when the friction term keeps a number of the conservation laws. 

Our ultimate objective is to use the PINN technology to discover the most appropriate governing equation of the GRS based on various observational data or phenomenological considerations. 
If we encounter a problem similar to those discussed in this paper, it could be resolved using some of the minor perturbations covered in this study.     

\vspace{0.5cm}

\noindent {\bf Acknowledgments} 
The authors would like to thank Satoshi Horihata, Hiroshi Kakuhata, Ryu Sasaki, Filip Blaschke,
and Pawe\l~Klimas for their practical advice and valuable comments. 
N.S. and K.S. would like to thank all the conference organizers of QTS12 and Prof.\v{C}estmir Burd\'{i}k for their hospitality and 
many kind considerations. 
We sincerely appreciate the anonymous reviewers' valuable comments and 
constructive suggestions, which have significantly improved the quality of 
this manuscript.
K.S. was supported by Tokyo University of Science. 
A.N., N.S., and K.T. were supported in part by JSPS KAKENHI Grant Number JP23K02794.

\vspace{0.5cm}

\appendix
\section*{Appendix A: Technical setup of the Physics-Informed Neural Networks}

\begin{figure}[h]
	\includegraphics[width=1.0\linewidth]{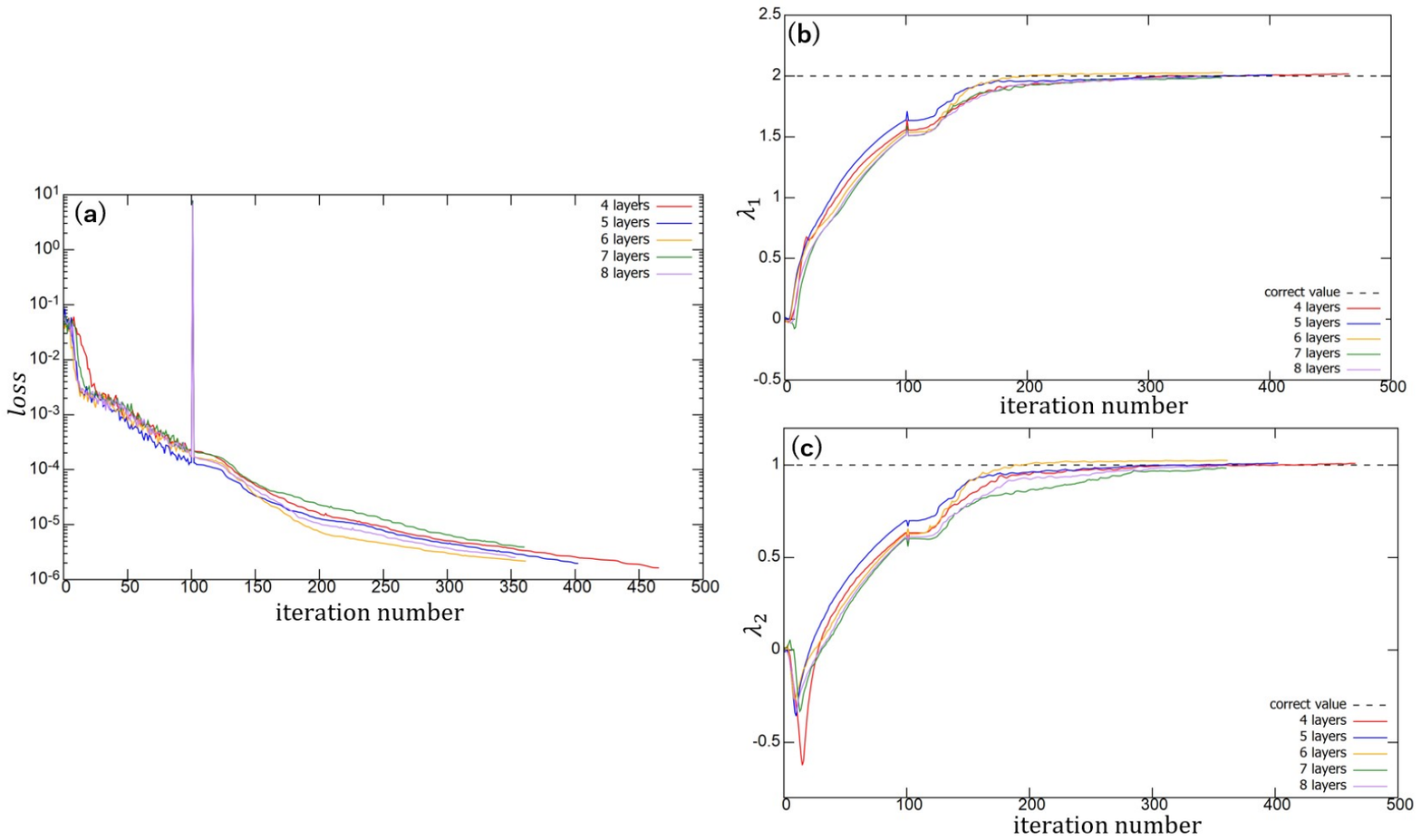}
	\caption{\label{ZK_setup1}The convergence of the loss function and the parameters of the ZK equation:  
	(a) the \textit{MSE}, (b) the predicted parameter of $\lambda_1$, (c) the predicted parameter of $\lambda_2$.}
\end{figure}

Our analysis of the PINNs in this paper is based on a neural network consisting 
of 4 hidden layers, each containing 20 nodes. 
In order to fix the optimal parameters for both layers and nodes, here 
we solve the inverse problem of the ZK equation
\begin{equation}
	u_t+\lambda_1uu_x+\lambda_2\left(\nabla^2 u\right)_x=0
\end{equation}
for changing the number of layers or nodes. 
In Fig.\ref{ZK_setup1}, we study the convergence property of the PINNs from $4$ to $8$ layers. 
This indicates that the result of the 6 layers is the best choice for convergence. 
Table \ref{table_ZK_setup1} is the ability of the parameter prediction where the 4 layer looks better than the 
other number of the layers. The findings imply that selecting the layers is not as crucial as one may 
think to achieve better solutions. 

\begin{table}[h]
	\begin{center}
		\caption{\label{table_ZK_setup1}The accuracy of the parameter prediction.}
		\begin{tabular}{ccc}\hline \hline
			layers&$\lambda_1$&$\lambda_2$\\ \hline
			4&2.01547&1.00873\\
			5&2.00811&1.01046\\
			6&2.02570&1.02479\\
			7&1.98557&0.98272\\
			8&2.00094&0.99105\\\hline\hline
		\end{tabular}
	\end{center}
\end{table}

Next, we study the appropriate size of the training data in the inverse analysis. 
We employ the range of the coordinates set 
as $x\in[-20,20],\ y\in[-10,10],\ t\in[0.0,5.0]$ and the mesh number 
is set as $N_x=200,\ N_y=200,\ N_t=1000$. 
The whole number of the training data becomes $4\times10^7$ in this setup. 
Due to computational resource limits, we cannot use the whole data set for the 
PINN analysis; instead, we are forced to select the data's collocation points at random. 
Surprisingly, Table \ref{appendix_points} indicates that the PINNs do not lose their predictability 
even if they learn only 0.0125\% of its whole data. In this paper, we always use 25000 points (0.0625\% of its whole data) as the training data.

\begin{table}[H]
	\begin{center}
		\caption{\label{appendix_points}The predictability for the parameters $\lambda_1,\lambda_2$ of the inverse analysis
		 for the choice of the training data points (and the ratio to the whole data).}
		\begin{tabular}{lcc}\hline \hline
			training points~(ratio) & $\lambda_1$&$\lambda_2$\\ \hline
			5000~(0.0125\%)&  2.00185&0.99356\\
			10000~(0.025\%)& 2.01124&1.01236\\
			15000~(0.0375\%)& 2.01065&1.00369\\
			20000~(0.05\%)& 1.99789&1.00730\\
			25000~(0.0625\%) & 2.01291&1.01996\\
			30000~(0.075\%)& 2.00883&1.00784\\
			35000~(0.0875\%)& 2.01950&1.01835
			\\ \hline\hline
		\end{tabular}
	\end{center}
\end{table}

\section*{References}

\bibliography{NN}

\end{document}